\documentclass[a4paper,11pt]{article}
\usepackage{jheppub} 
\usepackage{lineno}
\usepackage{xcolor}
\usepackage{tikz}
\usepackage{subfigure}
\usepackage{float}
\usetikzlibrary{arrows.meta}
\usetikzlibrary{calc}
\usepackage{ytableau}
\usepackage{makecell}
\usepackage{longtable}
\usepackage{amsmath}
\usepackage{comment}
\DeclareUnicodeCharacter{2212}{\ensuremath{-}}

\newcommand{\Qzz}{Q_{0,0}}
\newcommand{\Qzo}{Q_{0,1}}
\newcommand{\Qoz}{Q_{1,0}}
\newcommand{\Qzt}{Q_{0,2}}
\newcommand{\Qot}{Q_{1,2}}
\newcommand{\Qoo}{Q_{1,1}}
\newcommand{\ri}{\mathrm{i}}
\ytableausetup{mathmode, boxframe=normal, boxsize=2em}

\title{\boldmath Rational $Q$-systems for integrable spin chains \\ without $U(1)$ symmetry}

\author[a,c]{Yunfeng Jiang,}
\author[a]{Yi-Chao Liu,}
\author[b]{Yuan Miao,}
\author[a]{and Zi-Xi Tan}
\affiliation[a]{Shing-Tung Yau Center and School of Physics, Southeast University,Nanjing 210096, China}
\affiliation[b]{Kavli Institute for the Physics and Mathematics of the Universe (WPI), UTIAS, The University of Tokyo, Kashiwa, Chiba 277-8583, Japan}
\affiliation[c]{Peng Huanwu Center for Fundamental Theory, Hefei, Anhui 230026, China}

\emailAdd{jinagyf2008@seu.edu.cn}
\emailAdd{iqna@seu.edu.cn}
\emailAdd{yuan.miao@ipmu.jp} 
\emailAdd{zxtan@seu.edu.cn}

\abstract{The $Q$-system is an efficient method for finding complete physical solutions of Bethe ansatz equations, but so far its application has been confined to systems possessing $U(1)$ symmetry. We extend the rational $Q$-system framework to integrable spin chains without $U(1)$ symmetry, exemplified by the closed XXZ model with anti-diagonal twists and the open XXZ model with non-diagonal boundary fields. We demonstrate that the $Q$-system can be derived by combining $TQ$-relation with fusion relations of higher-spin transfer matrices. This yields $QQ$-relations analogous to the $U(1)$ symmetric case but incorporating additional inhomogeneous terms. We present numerical solutions that are validated against exact diagonalization, confirming that it generates all and exclusively physical solutions.}

\begin{document}
\maketitle
\flushbottom

\section{Introduction}
\label{sec:intro}

Solving the Bethe ansatz equations (BAE) is a key step in many applications of Bethe ansatz solvable quantum integrable models. Meanwhile, this is generally a non-trivial task. The situation differs depending on whether one wants to find certain specific solutions or all solutions. For specific solutions, such as those corresponding to the ground state and the first few excited states, one can readily apply numerical iterations based on reasonable assumptions about the distribution of Bethe roots. But the situation is quite different if the task is to find all solutions of the BAE for given quantum numbers.

To keep our discussions concrete and easier to follow, we take the BAE of the Heisenberg XXX$_{1/2}$ spin chain with periodic boundary condition as an example:
\begin{equation}
   \left(\frac{u_j + \frac{\ri}{2}}{u_j - \frac{\ri}{2}}\right)^L = \prod_{k \neq j}^{M} \frac{u_j - u_k + \ri}{u_j - u_k - \ri} \; , \quad j = 1 , 2 , \cdots M \;,
   \label{eq:BetheLM}
\end{equation}
where $L$ and $M$ are the length and magnon number, respectively. For fixed $L$ and $M$, the task is to find all physical solutions to BAE \eqref{eq:BetheLM}.\par

It is known that not all solutions of the BAE are physical. That is, some solutions do not yield eigenstates of the Hamiltonian when substituted into the Bethe wave function. Such solutions are called \emph{unphysical} and should be discarded. A better formulation is the $TQ$-relation \cite{Baxter1973-i,Baxter1973-ii,Baxter1973-iii,baxter1985exactly} , where instead of finding Bethe roots, one finds the polynomial $Q(u)$ by solving the $TQ$-relation. Then the zeroes of the $Q$-polynomial correspond to the Bethe roots. The $TQ$-relation is a better formulation because it already eliminates some unphysical solutions (such as repeated roots in the current example) and avoids possible divergences of the BAE. However, the $TQ$-relation alone does not eliminate all unphysical solutions. Finding additional constraints that accomplish this is an important and interesting question.\par 

The additional requirement has been found and can be formulated in various forms. In \cite{Hao_2013} , by performing a careful regularization of the wave function, Nepomechie obtained an additional requirement and numerically verified that it indeed yields all and only physical solutions. In 2016, Marboe and Volin \cite{Marboe:2016yyn} proposed the rational $Q$-system as a fast solver of the BAE. In addition to computational efficiency, the main merit of the $Q$-system approach is that it provides only the physical solutions. In other words, the condition found by Nepomechie is automatically encoded in the $Q$-system formalism in a certain way. This was demonstrated by Granet and Jacobsen in \cite{Granet:2019knz} . In the same paper, they also showed that the additional requirement is equivalent to requiring the other solution of the $TQ$-relation to be a polynomial of order $(L-M+1)$. It has been proven \cite{2007arXiv0706.0688M} that if both solutions of the $TQ$-relation are polynomials, the solution of the $TQ$-relation gives the complete spectrum of the Heisenberg XXX$_{1/2}$ spin chain. This proves that the $Q$-system indeed gives the complete spectrum of the XXX$_{1/2}$ chain.\par

The construction of Volin and Marboe has been generalized to more general models in the past years, including the $q$-deformed XXZ spin chain (both with generic $q$ \cite{Bajnok:2019zub} and $q$ at a root of unity \cite{Hou:2023jkr}) , higher spin models \cite{Hou:2023ndn,He:2025}, general $A_n$-type models \cite{Gu:2022dac,Nepomechie:2020ixi}, models with diagonal twist \cite{Bohm:2022ata} and parallel boundary magnetic fields \cite{Nepomechie:2019gqt}. All of the above models preserve at least the $U(1)$ symmetry of the global $SU(2)$ symmetry of the XXX$_{1/2}$ spin chain, \emph{i.e.}, the magnon numbers are conserved. Meanwhile, there are many interesting integrable models without $U(1)$ symmetry. The BAE of such models are typically much more challenging to solve. Given the developments in the $Q$-system, it is therefore natural to ask whether we can construct a $Q$-system like formalism for $U(1)$-breaking models. Here we distinguish two cases of $U(1)$ breaking. In the first case, the symmetry is broken by bulk interactions, such as in the XYZ spin chain. In the second case, the symmetry is broken by boundary conditions, such as the spin torus (closed chain with an anti-diagonal twist) \cite{Kulish2009} and the open chain with non-parallel boundary magnetic fields \cite{cao2015exact}. In this work, we construct the $Q$-system for the latter case.\par

There are two main results of this paper. The first concerns a new perspective on the $Q$-system. We find that the aforementioned additional requirement for the $TQ$-relation can also be obtained from the fusion relation of the transfer matrix. We derive the rational $Q$-system by combining the usual 
$TQ$-relation with the fusion relations of the transfer matrices. This observation paves the way for a natural generalization to a broad class of models, including $U(1)$-breaking ones. The reason is that the $TQ$-relation and the fusion relations are known for such models, although the explicit form of the $TQ$-relations differs from the $U(1)$-preserving cases.\par

Based on this observation, we derive the $Q$-system for two typical models that break $U(1)$ symmetry: the first is the XXZ spin chain with an anti-diagonal twist, and the second is the open XXZ spin chain with non-parallel boundary conditions. Compared to the $U(1)$-preserving case, the $QQ$-relations acquire inhomogeneous terms, similar to and inherited from the inhomogeneous $TQ$-relations in these cases. We solve the $Q$-system numerically and verify that it indeed yields all physical solutions. 

The rest of the paper is organized as follows. In Section~\ref{sec:TQfusion}, we revisit the $Q$-system for the Heisenberg XXZ$_{1/2}$ spin chain with periodic boundary conditions and demonstrate that it can be derived from the $TQ$-relation and the fusion relations of the transfer matrices. In Section~\ref{sec:antidiagonal}, we generalize this construction to the Heisenberg spin chain with an anti-diagonal twist. Section~\ref{sec:unparallel} examines the open chain with non-diagonal boundary conditions. We present our conclusions and discuss future research directions in Section~\ref{sec:concl}. Detailed derivations and proofs of some of the statements in the main text are provided in the appendices.

\section{\texorpdfstring{$Q$-system from $TQ$- and fusion}{Q-system from TQ- and fusion}}
\label{sec:TQfusion}

In this section, we revisit the $Q$-system of XXZ spin chain with periodic boundary condition~\cite{Marboe:2016yyn, Nepomechie:2019gqt}. We will show the equivalence between the $Q$-system and the $TQ$- together with fusion relations. This serves as the foundation for establishing the $Q$-systems for models without $U(1)$ symmetry in subsequent sections.

The Hamiltonian of the XXZ spin chain with periodic boundary conditions is given by~\cite{PhysRev.112.309}
\begin{equation}
    H=\sum^L_{j=1}\left[\sigma_j^x\sigma_{j+1}^x+\sigma_j^y\sigma_{j+1}^y+ \Delta \, \sigma_j^z\sigma_{j+1}^z\right] \;, \quad\quad \vec{\sigma}_{L+1}\equiv\vec{\sigma}_1 \;,
    \label{eq:HXXZPBC}
\end{equation}
where $\vec{\sigma}_j=(\sigma^x_j,\sigma^y_j,\sigma^z_j)$ are Pauli matrices acting non-trivially on the $j$-th site of the total Hilbert space $\mathcal{H}_L =(\mathbb{C}^2)^{\otimes L}$ and $\Delta=\cosh\eta$ is the anisotropy. We will also use the parametrization $q=e^\eta$ such that $\Delta = \frac{q+q^{-1}}{2}$. In this work, we assume that $q$ is a generic number, that means that it is not at any root of unity value.

\subsection{\texorpdfstring{$TQ$-relation}{TQ-relation}}
The XXZ spin chain is Yang--Baxter integrable with the 6-vertex $R$-matrix \cite{baxter1972one, baxter1985exactly, Gaudin_2014, korepin1997quantum}. 
\begin{equation}
    \mathbf{R}_{m,n} (u,\eta) = \begin{pmatrix}
        \sinh(u+\eta) & {\color{gray} 0} & {\color{gray} 0} & {\color{gray} 0} \\
        {\color{gray} 0} & \sinh u & \sinh \eta & {\color{gray} 0} \\
        {\color{gray} 0} & \sinh \eta & \sinh u & {\color{gray} 0} \\
        {\color{gray} 0} & {\color{gray} 0} & {\color{gray} 0} & \sinh(u+\eta)
    \end{pmatrix}_{m,n} \; .
\label{eq:R_mat_XXZ}
\end{equation}
It satisfies the Yang-Baxter equation~\cite{baxter1985exactly, Gaudin_2014, korepin1997quantum}
\begin{equation}
    \mathbf{R}_{12}(u-v) \mathbf{R}_{13}(u) \mathbf{R}_{23}(v)= \mathbf{R}_{23}(v) \mathbf{R}_{13}(u) \mathbf{R}_{12}(u-v) \;.
\end{equation}
The monodromy matrix $\mathbf{M}_a(u)$ and the transfer matrix $\mathbf{T}(u)$ are constructed as~\cite{sklyanin1992quantum, faddeev1996algebraicbetheansatzworks, korepin1997quantum}~\footnote{We shift the argument $u$ in the transfer matrix for later convenience.}
\begin{equation}
    \mathbf{M}_a(u)=\mathbf{R}_{a1}(u)\mathbf{R}_{a2}(u) \cdots \mathbf{R}_{aL}(u) \;, \quad \mathbf{T}(u)=\text{tr}_a\left[\mathbf{M}_a(u-\eta/2)\right] \; ,
    \label{eq:monodromy_and_transfer}
\end{equation}
where $a$ denotes the auxiliary space.
Using the Yang-Baxter equation, one can show that the transfer matrices commute with each other
\begin{equation}
    \left[\mathbf{T}(u),\mathbf{T}(v)\right]=0 \; .
\end{equation}
Expanding $\log\mathbf{T}(u)$ in terms of the spectral parameter $u$, the above relation implies the existence of extensively many local conserved charges, which is a hallmark of quantum integrability.

Taking the auxiliary space to be $\mathbb{C}^2$, the monodromy and transfer matrices can be written as
\begin{equation}
    \mathbf{M}_a(u)=\begin{pmatrix}
        \mathbf{A}(u) & \mathbf{B}(u)\\ \mathbf{C}(u) &\mathbf{D}(u)
    \end{pmatrix}_a \;, \quad \mathbf{T}(u)=\mathbf{A}(u-\eta/2)+\mathbf{D}(u-\eta/2) \;.
 \label{eq:MandTPBC}
\end{equation}
The transfer matrix can be diagonalized by the algebraic Bethe ansatz. To this end, we start with the ferromagnetic vacuum state $| 0 \rangle = |\uparrow \uparrow \cdots \uparrow \rangle =\begin{pmatrix}
        1\\0
    \end{pmatrix}^{\otimes L}$, which satisfies the following relations
\begin{equation}
    \mathbf{A}(u)|0\rangle=a(u)|0\rangle\;, \quad \mathbf{D}(u)|0\rangle=d(u)|0\rangle\;, \quad \mathbf{C}(u)|0\rangle=0 \;.
\end{equation}
Here $a(u)=\sinh^L(u+\eta)$ and $d(u)=\sinh^L(u)$. The off-diagonal elements $\mathbf{B}(u)$ and $\mathbf{C}(u)$ play the role of creation and annihilation operators respectively. The Bethe states are obtained by acting $\mathbf{B}$-operators on the vacuum state $|0\rangle$
\begin{equation}
    | \{ u_j \}_{j=1}^M \rangle\equiv\mathbf{B}(u_1)\dots\mathbf{B}(u_M)|0\rangle \; .
\end{equation}

This state is an eigenstate of the transfer matrix $\mathbf{T}(u)$ with eigenvalue $T(u)$, \emph{i.e.}
\begin{equation}
    \mathbf{T}(u)|\{ u_j \}_{j=1}^M\rangle=T(u)|\{ u_j \}_{j=1}^M\rangle \;,
\end{equation}
if the Bethe roots $\{ u_j \}_{j=1}^M$ satisfy the Bethe ansatz equation~(BAE)
\begin{equation}
\label{eq:BAEXXZ}
 \left(\frac{\sinh(u_j+\frac{\eta}{2})}{\sinh(u_j-\frac{\eta}{2})}\right)^L=\prod^M_{k=1\atop k\neq j}\frac{\sinh(u_j-u_k+\eta)}{\sinh(u_j-u_k-\eta)} \;, \quad j\in \{1,2,...,M\} \;.
\end{equation}
The BAE can be written equivalently as the following $TQ$-relation
\begin{equation}
\label{eq:TQPBC}
 T(u)Q(u)=\sinh^L\left(u+\frac{\eta}{2}\right)Q(u-\eta)+\sinh^L\left(u-\frac{\eta}{2}\right)Q(u+\eta) \; ,
\end{equation}
with the $Q$-function
\begin{equation}
    Q(u)=\prod^M_{j=1}\sinh(u-u_j)\,.
\end{equation}
By construction, the zeros of the $Q$-function are the Bethe roots. To derive BAE from the $TQ$-relation, we take the limit $u \to u_j$ in \eqref{eq:TQPBC}. Since the left-hand side vanishes in this limit, the $TQ$-relation becomes \eqref{eq:BAEXXZ}.

\subsection{Fusion relations}\label{sec:fusion}
In addition to the transfer matrix defined in \eqref{eq:MandTPBC}, the higher-spin transfer matrices $\mathbf{T}_s(u)$ of the XXZ spin chain also play a crucial role in our construction. Following the notation in \cite{Miao_2021}, $\mathbf{T}_s(u)$ denotes the transfer matrix constructed from a monodromy matrix whose auxiliary space is in the spin-$s$ representation of $\mathcal{U}_q(\mathfrak{sl}_2)$. In particular, the transfer matrix in \eqref{eq:MandTPBC} is $\mathbf{T}(u) = \mathbf{T}_{1/2}(u)$. To construct the higher-spin transfer matrices, one can use the fusion procedure~\cite{kulish1981yang, Krichever1997, klumper1992conformal, Wiegmann_1997}, which leads to a recursive relation between different higher-spin transfer matrices. These relations are known as fusion relations and are given by~\cite{Krichever1997, Miao_2021}
\begin{equation}
    \mathbf{T}_{1/2}^{[(2s+1)\pm]}\left( u\right) \mathbf{T}_s(u)=T_0^{[(2s+2)\pm]}\left(u\right)\mathbf{T}_{s-1/2}^{[\mp]}( u)+T_0 ^{[2s\pm]}(u) \mathbf{T}_{s+1/2}^{[\pm]}\left( u\right) \;,
    \label{eq:fusion}
\end{equation}
where $T_0(u)=\sinh^L(u)$ and we have introduced the following shorthand notations
\begin{align}
f^{[a]}(u)\equiv f\left(u+a\tfrac{\eta}{2}\right),\qquad f^{[a]}(t)\equiv f\left(t q^{a/2}\right)\,,
\end{align}
and $f^{[\pm]}(u)\equiv f(u\pm\tfrac{\eta}{2})$, $f^{[\pm]}(t)\equiv f(tq^{\pm1/2})$ with $t=e^u$.
Note that \eqref{eq:fusion} is an operator equation. Working with the eigenvalues of $\mathbf{T}_{s}(u)$, denoted as $T_s(u)$, we obtain a functional equation of the same form
\begin{equation}
   {T}_{1/2}^{[(2s+1)\pm]}\left( u\right){T}_s(u)=T_0^{[(2s+2)\pm]}\left(u\right){T}_{s-1/2}^{[\mp]}( u)+T_0 ^{[2s\pm]}(u){T}_{s+1/2}^{[\pm]}\left( u\right) \;.
 \label{eq:fusionfunc}
\end{equation}

\paragraph{The `$TQ$+fusion' system} We combine the $TQ$-relation \eqref{eq:TQPBC} and the fusion relation \eqref{eq:fusionfunc} and call the combined system the `$TQ$+fusion' system
\begin{align}
&T_{1/2}(u)Q(u)=T_0^{[+]}(u)Q(u-\eta)+T_0^{[-]}(u)Q(u+\eta)\,,\\\nonumber
&{T}_{1/2}^{[(2s+1)\pm]}\left( u\right){T}_s(u)=T_0^{[(2s+2)\pm]}\left(u\right){T}_{s-1/2}^{[\mp]}( u)+T_0 ^{[2s\pm]}(u){T}_{s+1/2}^{[\mp]}\left( u\right)\,.
\end{align}
By requiring $Q(u)$ and all $T_s(u)$ to be Laurent polynomials in $t=e^u$, we can solve for $Q(u)$ and find the Bethe roots. It can be proven that if $T_{1/2}(u)$ and $T_1(u)$ are Laurent polynomials in $t$, then all $T_s(u)$ ($s=\frac{3}{2},1,\frac{5}{2},\cdots$) are Laurent polynomials in $t$ (see Appendix~\ref{app:recursion} for a proof). This implies that it is sufficient to consider $s=1/2$ in the fusion relation. Written in terms of multiplicative variables, we consider the following system
\begin{align}
&T_{1/2}(t)Q(t)=T_0^{[+]}(t)Q^{[2-]}(t)+T_0^{[-]}(t)Q^{[2+]}(t)\,,\label{eq:TQfusion1}\\
&{T}_{1/2}^{[2\pm]}\left(t\right){T}_{1/2}(t)=T_0^{[3\pm]}\left(t\right){T}_{0}^{[\mp]}( t)+T_0 ^{[\pm]}(t){T}_{1}^{[\pm]}\left( t\right)\,.\label{eq:TQfusion2}
\end{align}
To solve the above system, we first make a Laurent polynomial ansatz for $Q(t)$ 
\begin{align}
Q(t)=\frac{1}{2^M}\left(t^M+\frac{t^{-M}}{c_0}+\sum_{k=1}^{M-1}c_k\,t^{2k-M}\right)\,,
\end{align}
with undetermined coefficients $\{c_k\}$. From the $TQ$-relation \eqref{eq:TQfusion1}, we obtain
\begin{align}
\label{eq:expressT12}
T_{1/2}(t)=\frac{T_0^{[+]}(t)Q^{[2-]}(t)+T_0^{[-]}(t)Q^{[2+]}(t)}{Q(t)}\,.
\end{align}
To ensure that $T_{1/2}(t)$ is a Laurent polynomial in $t$, we need to require the remainder of the above quotient is zero, which gives a set of algebraic equations for $\{c_k\}$. Using the fusion relation \eqref{eq:TQfusion2}, we obtain $T_1(t)$
\begin{align}
T_1(t)=\frac{T_{1/2}^{+}(t)T_{1/2}^-(t) - T_0^{[2+]}(t)T^{[2-]}_0(t)}{T_0(t)}\,.
\end{align}
Plugging \eqref{eq:expressT12} in the above relation and requiring the remainder of the quotient to be zero gives another set of algebraic equations for $\{c_k\}$. The two sets of algebraic equations can be solved numerically or analytically, resulting in $Q(t)$.

\subsection{\texorpdfstring{$Q$-system}{Q-system}}

\begin{figure}[H]
    \centering
\begin{tikzpicture}[scale=1.0,
    box/.style={draw, minimum size=1.2cm},
    arrow/.style={->, >=stealth, thick},
    dot/.style={circle, fill=black, inner sep=0pt, minimum size=3pt}]
    \foreach \x in {0,1,2} {
        \node[box] at (\x*1.2, 0) {};
    }
    \foreach \x in {0,1,2,3,4,5} {
        \node[box] at (\x*1.2, -1.2) {};
    }
    \draw[arrow] (-0.8,0) -- (-0.8,1);
    \node at (-0.8, 1.2) {$a$};
    \draw[arrow] (6,-2) -- (7,-2);
    \node at (7.2,-2) {$b$};
    \draw[arrow] (7,0) -- (8,0);
    \node at (9,0) {$M$ boxes};
    \draw[arrow] (7,-1.2) -- (8,-1.2);
    \node at (9.4,-1.2) {$L-M$ boxes};
    \node[fill=black, circle, inner sep=1pt] at (-0.5*1.2,-1.5*1.2) {};
    \node[fill=black, circle, inner sep=1pt] at (-0.5*1.2,-0.5*1.2) {};
    \node[fill=black, circle, inner sep=1pt] at (2.5*1.2,0.5*1.2) {};
    \node at (-0.85*1.2,-1.5*1.2) {$Q_{0,0}$};
    \node at (-0.85*1.2,-0.5*1.2) {$Q_{1,0}$};
    \node at (2.5*1.2,0.8*1.2) {$Q_{2,3}$};
    \end{tikzpicture}
    \caption{The Young diagram corresponding to the rational $Q$-system with $L=9$ and $M=3$. The subscript of the notation $Q_{a,b}$ denotes its coordinate on the lattice: $(a,b)$ denotes the site which is located at the $a$-th row from the bottom to the top and the $b$-th column from the left to the right. }
    \label{fig:Young_diagram}
\end{figure}
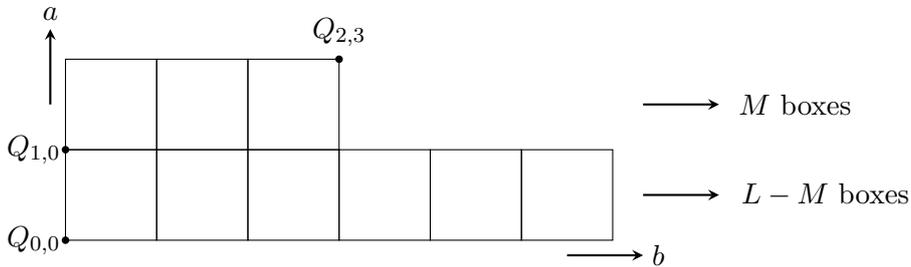

Now we review the rational $Q$-system. For the $Q$-system corresponding to the BAE with length $L$ and $M$ magnons, we associate a Young diagram $(L-M,M)$, as is shown in Fig.~\ref{fig:Young_diagram}. We define a $Q$-function $Q_{a,b}(u)$ on each vertex $(a,b)$ of the Young diagram. These $Q$-functions are related to each other by the $QQ$-relations
\begin{equation}
Q_{a+1,b}Q_{a,b+1}= Q_{a+1,b+1}^{[+]}Q_{a,b}^{[-]}-Q_{a+1,b+1}^{[-]}Q_{a,b}^{[+]}\;.
\label{eq:QQPBC}
\end{equation}

By fixing the $Q$-functions at the left and upper boundaries and imposing proper analytic conditions, we can solve for all the $Q$-functions.
Specifically, we set $Q_{2,b}=1$ at the upper boundary and~\cite{Bajnok_2020}
\begin{equation}
\label{eq:leftBdr}
  \Qzz(u)=\sinh^L(u) \; , \qquad \Qoz(u)=\prod_{j=1}^M\sinh(u-u_j) \;,
\end{equation}
at the left boundary. Here $Q_{1,0}$ is identified with Baxter's $Q$-function whose zeros are the Bethe roots. We impose the condition that all $Q_{a,b}(t)$ are Laurent polynomials of $t$. Using the $QQ$-relations we can express all $Q_{a,b}$ in terms of $Q_{1,0}$. Requiring that they are Laurent polynomials lead to a set of algebraic equations for the coefficients of $Q_{1,0}$ called the zero remainder conditions (ZRC). Solving the ZRC gives $Q_{1,0}(t)$.

\paragraph{Minimal $Q$-system} For the spin-$\frac{1}{2}$ XXZ chain, it can be proven that one only needs to solve the ZRC of $Q_{0,1}$ and $Q_{0,2}$~\cite{hou2024rational}, the ZRC for the rest $Q$-functions are automatically satisfied. Therefore we only need to solve the following minimal $Q$-system
\begin{align}
\label{eq:minimalQsys}
&Q_{1,n}=Q_{1,n-1}^{[+]}-Q_{1,n-1}^{[-]}\,,\quad (n=1,2)\,,\\
&Q_{1,0}Q_{0,1}=Q^{[+]}_{1,1}Q_{0,0}^{[-]}-Q_{1,1}^{[-]}Q_{0,0}^{[+]}\,,\label{eq:QQ1}\\
&Q_{1,1}Q_{0,2}=Q_{1,2}^{[+]}Q_{0,1}^{[-]}-Q_{1,2}^{[-]}Q_{0,1}^{[+]}\,.\label{eq:QQ2}
\end{align}

\subsection{Equivalence of two systems}\label{sec:EQTS}
Now we state the main result of this section.
\paragraph{Theorem 1} The `$TQ$+fusion' system \eqref{eq:TQfusion1} is equivalent to the minimal $Q$-system \eqref{eq:minimalQsys} by the following identification
\begin{align}
&Q_{0,0}=T_0\,,\qquad Q_{1,0}=Q\,,\label{eq:identify1}\\
&Q_{0,1}=T_{1/2}-\left(T_0^{[+]}+T_0^{[-]}\right)\,,\label{eq:identify2}\\
&Q_{0,2}=T_1-2\left( T_{1/2}^{[+]}+T_{1/2}^{[-]}\right)+\left(T_0^{[2+]}+T_0^{[2-]}+3T_0\right)\,\label{eq:identify3}.
\end{align}
\noindent\emph{Proof} We will start from the $TQ$+fusion system and derive the minimal $Q$-system.
The identification rules 
\begin{align}
\label{eq:id1}
Q_{0,0}=T_0=\sinh^L(u),\qquad Q_{1,0}=Q
\end{align}
are trivial. We then define the $Q$-functions $Q_{1,n}=Q_{1,n-1}^{[+]}-Q_{1,n-1}^{[-]}$ in terms of $Q$. In particular,
\begin{align}
\label{eq:id2}
&Q_{1,1}=Q_{1,0}^{[+]}-Q_{1,0}^{[-]}=Q^{[+]}-Q^{[-]}\,,\\\nonumber
&Q_{1,2}=Q_{1,1}^{[+]}-Q_{1,1}^{[-]}=Q^{[2+]}-2Q+Q^{[2-]}\,.
\end{align}
We first show that the $TQ$-relation is equivalent to the $QQ$-relation \eqref{eq:QQPBC} for $(a,b)=(0,0)$. Using the identification rules \eqref{eq:id1} and \eqref{eq:id2}, the right-hand side of the $QQ$-relation \eqref{eq:QQ1} can be written as
\begin{align}
Q_{1,1}^{[+]}Q_{0,0}^{[-]} - Q_{1,1}^{[-]} Q_{0,0}^{[+]} =&\,(Q^{[2+]}-Q)T_0^{[-]}-(Q-Q^{[2-]})T_0^{[+]}\\\nonumber
=&\,\big(T_0^{[+]}Q^{[2-]}+T_0^{[-]}Q^{[2+]}\big)-Q(T_0^{[+]}+T_0^{[-]})\\\nonumber
=&\,\big(T_{1/2}-(T_0^{[+]}+T_0^{[-]})\big)Q,
\end{align}
where in the second line we have used the $TQ$-relation. Using the identification \eqref{eq:identify2}, we obtain exactly the left-hand side of the $QQ$-relation \eqref{eq:QQ1}.\par

To derive the last $QQ$-relation \eqref{eq:QQ2}, we need the following relation
\begin{align}
\label{eq:intermediate}
T_1Q_{1,1}=\big(T_0^{[2+]}-T_{1/2}^{[+]}\big)Q_{1,0}^{[3-]}-\big(T_0^{[2-]}-T_{1/2}^{[-]}\big)Q_{1,0}^{[3+]}.
\end{align}
This can be proven as follow: multiply both side of the fusion relation \eqref{eq:TQfusion2} by $Q_{1,1}$, we obtain
\begin{align}
T_0T_1Q_{1,1}=T_{1/2}^{[+]}T_{1/2}^{[-]}Q_{1,1}-T_0^{[2+]}T_0^{[2-]}Q_{1,1}\,.
\end{align}
Plugging $Q_{1,1}=Q^{[+]}-Q^{[-]}$ on the right-hand side of the above relation and using the $TQ$-relation recursively, we obtain
\begin{equation}
    T_0 T_1 Q_{1,1} = T_0\left(\left(T_0^{[2+]}-T_{1/2}^{[+]}\right)\Qoz^{[3-]}-\left(T_0^{[2-]}-T_{1/2}^{[-]}\right)\Qoz^{[3+]}\right) \; ,
\end{equation}
Dividing both sides by $T_0$ leads to \eqref{eq:intermediate}. \par

Now we move on to derive the $QQ$-relation \eqref{eq:QQ2}. The right-hand side of the $QQ$-relation can be written as
\begin{align}
\label{eq:rhsQQ01}
Q_{1,2}^{[+]}Q_{0,1}^{[-]}-Q_{1,2}^{[-]}Q_{0,1}^{[+]}=&\,\big(Q_{1,1}^{[2+]}-Q_{1,1}\big)Q_{0,1}^{[-]}-\big(Q_{1,1}-Q_{1,1}^{[2-]}\big)Q_{0,1}^{[+]}\\\nonumber
=&\,Q_{0,1}^{[-]}Q_{1,1}^{[2+]}+Q_{0,1}^{[+]}Q_{1,1}^{[2-]}-Q_{1,1}\big(Q_{0,1}^{[-]}+Q_{0,1}^{[+]}\big)\,.
\end{align}
Using the relation \eqref{eq:intermediate}, one can show that
\begin{align}
Q_{0,1}^{[-]}Q_{1,1}^{[2+]}+Q_{0,1}^{[+]}Q_{1,1}^{[2-]}=\left(T_1+\Qzz-T_{1/2}^{[+]}-T_{1/2}^{[-]}\right)\Qoo\,.
\end{align}
Therefore, by identifying
\begin{align}
Q_{0,2}=&\,T_1+Q_{0,0}-\big(T_{1/2}^{[+]}-T_{1/2}^{[-]}\big)-\big(Q_{0,1}^{[-]}+Q_{0,1}^{[+]}\big)\\\nonumber
=&\,T_1-2\big(T_{1/2}^{[+]}-T_{1/2}^{[-]}\big)+\left(T_0^{[2+]}+T_0^{[2-]}+3T_0 \right)\,,
\end{align}
\eqref{eq:rhsQQ01} can be written as
\begin{align}
Q_{1,1}Q_{0,2}=Q_{1,2}^{[+]}Q_{0,1}^{[-]}-Q_{1,2}^{[-]}Q_{0,1}^{[+]}\,,
\end{align}
which is precisely the $QQ$-relation we are after. Therefore the theorem is proven.

The key lesson we learned from the theorem is that the $Q$-system can be derived from combining the $TQ$-relation and the fusion relations, which provides another perspective of the $Q$-system. More importantly, we can take this equivalence as a \emph{guiding principle} to construct $Q$-systems. The main advantage of this point of view is that for most (if not all) Bethe ansatz solvable models the $TQ$-relations are known and the fusion relations can be derived once the $R$-matrix is given. Therefore, it allows us to construct the $Q$-system for very general models, including the models without $U(1)$ symmetry.

\section{XXZ chain with anti-diagonal twist}
\label{sec:antidiagonal}

In this section, we construct the $Q$-system for XXZ spin chain with an anti-diagonal twist, which explicitly breaks the $U(1)$ symmetry generated by $S^z=(\sigma_1^z+\cdots+\sigma_L^z)/2$.

\subsection{Review of diagonal twist}

The $Q$-system for the XXZ spin chain with a diagonal twist is known in \cite{Bajnok_2020}, where the $U(1)$ symmetry is preserved. The twisted boundary conditions can be regarded as a generalization of the periodic boundary, which only changes the periodic boundary condition $\sigma^\alpha_{L+1} = \sigma^\alpha_1$ into a unitary transformation of the boundary spin operator under a twist matrix $\mathbf{W}$, namely $\sigma^\alpha_{L+1} = \mathbf{W} \sigma^\alpha_1 \mathbf{W}^{-1}$. Taking the diagonal twist matrix $\mathbf{W} = \mathrm{diag}(1, e^{\mathrm{i}\theta})$, we have
\begin{equation}
    \sigma^\pm_{L+1} = e^{\pm\mathrm{i}\theta} \sigma^\pm_1, \qquad \sigma^z_{L+1} = \sigma^z_1\,.
\end{equation}
In this case, the $Q$-system is similar to \eqref{eq:QQPBC}, but with an additional twist factor \cite{He:2025},
\begin{equation}
    Q_{a+1,b}Q_{a,b+1}= Q_{a+1,b+1}^{[+]}Q_{a,b}^{[-]}-\kappa_a Q_{a+1,b+1}^{[-]}Q_{a,b}^{[+]}, \qquad \kappa_0 = e^{\mathrm{i}\theta}, \kappa_1=1.
\end{equation}
The boundary condition is exactly the same as the periodic boundary case \eqref{eq:leftBdr}, and the procedure for finding the Bethe roots is the same.

\subsection{Anti-diagonal twist}

With both periodic boundary condition and diagonal twist boundary condition, $U(1)$ symmetry is preserved, which allows one to solve the BAE in subspaces of fixed magnon numbers. However, for twist boundary conditions that break $U(1)$ symmetry, this is no longer the case. Therefore the situation is more complicated. The twist matrix $\mathbf{W}$ preserves integrability as long as it commutes with the $R$-matrix \cite{deVega:1984ig}, \emph{i.e.}
\begin{equation}
    \left[ \mathbf{R}_{mn}(u-v), \mathbf{W}_m \otimes \mathbf{W}_n \right] = 0.
\end{equation}
We consider a general anti-diagonal twist matrix
\begin{equation}
    \mathbf{W}_a = \begin{pmatrix}
        0 & \alpha \\
        \beta & 0 
    \end{pmatrix}_a \; ,
\end{equation}
and define the twisted monodromy matrix
\begin{equation}
    \mathbf{M}_{a}^{\rm AD} (u) = \mathbf{W}_a \mathbf{R}_{a1}(u) \mathbf{R}_{a2}(u) \cdots \mathbf{R}_{aL}(u) = \begin{pmatrix}
        \alpha \mathbf{C} (u) & \alpha \mathbf{D} (u) \\ \beta \mathbf{A}(u) & \beta \mathbf{B} (u) 
    \end{pmatrix}_a \,.
    \label{eq:Xtwistmonodromy}
\end{equation}
The transfer matrix is defined as 
\begin{equation}
    \mathbf{T}_{1/2}^{\rm AD} (u) = \mathrm{Tr}_a (\mathbf{M}_{a} (u-\eta/2)) \; .
    \label{eq:ThalfAD}
\end{equation}
The model is still integrable, meaning that the transfer matrices commute, \textit{i.e.}
\begin{equation}
    \left[ \mathbf{T}_{1/2}^{\rm AD} (u) , \mathbf{T}_{1/2}^{\rm AD} (v) \right] = 0 \; , \quad \forall u,v \in \mathbb{C} \; .
\end{equation}
The Hamiltonian can be obtained from the transfer matrix \eqref{eq:ThalfAD} by taking its logarithmic derivative as usual, the only difference is now we have $\sigma^\alpha_{L + 1} = \mathbf{W} \sigma^\alpha_1 \mathbf{W}^{-1}$, which leads to
\begin{equation}
\begin{split}
    \mathbf{H}^{\rm AD} (\alpha , \beta) = & \sum_{j=1}^{L-1} \left( \sigma^x_j \sigma^x_{j+1} + \sigma^y_j \sigma^y_{j+1} + \Delta \sigma^z_j \sigma^z_{j+1} \right) \\
    & + \sigma^x_L \left(p_1 \sigma^x_{1} + p_2 \sigma^y_{1}\right) + \sigma^y_L \left(p_2 \sigma^x_{1} - p_1 \sigma^y_{1}\right) - \Delta \sigma^z_L \sigma^z_{1} \; ,
\end{split}
\end{equation}
where 
\begin{equation}
p_1 =\frac{\alpha}{2\beta} + \frac{\beta}{2\alpha}, \qquad  p_2 = \ri \left(\frac{\alpha}{2\beta} - \frac{\beta}{2\alpha}\right).
\end{equation}
This Hamiltonian possesses a $\mathbb{Z}_2$ symmetry generated by the spin flip operator when $\alpha = \beta$,
\begin{equation}
    \left[ \mathbf{H}^{\rm AD} (\alpha, \alpha) , \prod_{n=1}^L \sigma^x_n \right] = 0 \; .\label{eq:HZ2}
\end{equation}
However, the spin flip $\mathbb{Z}_2$ operator does not commute with the transfer matrix \eqref{eq:ThalfAD}. We will see the consequence of this $\mathbb{Z}_2$ symmetry in subsection~\ref{subsec:ADnumeric}.

\paragraph{The `$TQ$ + Fusion' System} To derive the $Q$-system, we follow the same strategy by combining the $TQ$-relation and fusion relation. The $TQ$-relation for the model is known \cite{Cao_2015} and is given by
\begin{equation}
  q^{-\frac{1}{2}} T_{1/2} Q = \alpha t T^{[-]}_0 Q^{[2+]} - \beta t^{-1} T^{[+]}_0 Q^{[2-]} - \alpha \beta c(t) T^{[+]}_0 T^{[-]}_0 \; ,
  \label{eq:TQAD}
\end{equation}
where $T_0 = \frac{1}{2^L} (t-t^{-1})^{L}$ and $T_{1/2}$ is the eigenvalue of the transfer matrix $\mathbf{T}_{1/2}^{\rm AD}.$ In order not to be bothered with additional superscripts, we will slightly abuse the notation and use $T_s$ to denote the eigenvalue of the spin-$s$ transfer matrix with twisted boundary conditions. $T_s$ is a Laurent polynomial in $t=e^u$ of order $(L-1)$, and the $Q$-polynomial is a Laurent polynomial in $t$ of order $L$
\begin{equation}
    Q (t) = \prod_{j = 1}^L \sinh(u - u_j) = \frac{1}{2^L} \prod_{j=1}^L \left( \frac{t}{t_j} - \frac{t_j}{t} \right) = \frac{1}{2^L} \sum_{n=0}^L d_n t^{L-2n} \; .
\end{equation}
The function $c(t)$ in \eqref{eq:TQAD} is given by
\begin{align}
 \label{eq:ITct}
   c(t) = &\,2^L q^{L/2} \left( \beta^{-1} \frac{t}{\prod_{j=1}^L t_j} - \alpha^{-1} \frac{\prod_{j=1}^L t_j}{t} \right)\\\nonumber
     =&\, 2^L q^{L/2} \left( \beta^{-1} d_0 t - \alpha^{-1} (-1)^L d_L t^{-1} \right) \;,
\end{align}
where
\begin{equation}
    d_0 = \prod_{j=1}^L t_j^{-1} \; , \quad d_L = (-1)^L \prod_{j=1}^L t_j \; .
\end{equation}
The fusion relation is given by
\begin{equation}
    T_0 T_1 = T_{1/2}^{[+]} T_{1/2}^{[-]} + \alpha \beta T_0^{[2+]} T_0^{[2-]} \; .
\label{eq:fusionAD1}
\end{equation}
Combining \eqref{eq:TQAD} and \eqref{eq:fusionAD1}, we obtain the $TQ$+fusion system for this model.

\paragraph{Minimal Q-system}
We will demonstrate that the $TQ$+fusion system is equivalent to a minimal $Q$-system, with the following $QQ$-relations
\begin{subequations}
    \begin{align}
        Q_{1,1} &= t^{[-]} Q^{[+]}_{1,0} - (t^{-1})^{[+]} Q^{[-]}_{1,0} \; , \label{eq:Q11AD}\\
        Q_{1,2} &= t Q^{[+]}_{1,1} - t^{-1} Q^{[-]}_{1,1} \; , \label{eq:Q12AD}\\
        \Qzo \Qoz &= \beta \Qzz^{[+]} \Qoo^{[-]} + \alpha \Qzz^{[-]} \Qoo^{[+]} - \alpha \beta c(t) Q^{[+]}_{0,0}Q^{[-]}_{0,0} \; ,  \label{eq:Qsystem1stAD}\\
        \Qzt \Qoo &= \beta \Qzo^{[+]} \Qot^{[-]} + \alpha \Qzo^{[-]} \Qot^{[+]} - F(t) \; ,  \label{eq:Qsystem2ndAD}\\
        F(t) &= \alpha \beta t^{[+]} (c(t))^{[+]} \Qzo^{[-]} \Qzz^{[2+]} - \alpha \beta (t^{-1})^{[-]} (c(t))^{[-]} \Qzo^{[+]} \Qzz^{[2-]} \; .\notag
    \end{align}\label{eq:QSAT}
\end{subequations}
where the definition of $c(t)$ has been given in \eqref{eq:ITct}.  The additional factors of $t$ and $t^{-1}$ are introduced for $Q_{1,n}$ in order to obtain more compact $QQ$-relations. As is shown in Appendix \ref{app:QQ_relation_2nd_box}, the form of $QQ$-relation is not unique. Without the prefactors, the $QQ$-relations will take a more complicated form. Analogous constructions will also be used in the open boundary case.

We claim that if we define the identification rules as follows, the minimal $Q$-system \eqref{eq:QSAT} is equivalent to previous `$TQ$ + Fusion' system~\eqref{eq:TQAD} and \eqref{eq:fusionAD1},
\begin{subequations}
    \begin{align}
    Q_{0,0} &= T_0 \; , \,\, Q_{1,0} = Q \; , \label{eq:IQ0Q1}\\
    Q_{0,1} &= q^{-\frac{1}{2}}T_{1/2} + q^{-1}\left(\beta t T_0^{[+]} - t^{-1} \alpha T_0^{[-]}\right) \; , 
    \label{eq:Q01AD}\\
    \Qzt &= T_1 + q^{-1/2} \left(\beta t \Qzo^{[+]} - \alpha t^{-1} \Qzo^{[-]}\right) + q^{-1} \left(\beta t T_{1/2}^{[+]} - \alpha t^{-1} T_{1/2}^{[-]}\right) - q^{-2} \alpha \beta T_0\;.\label{eq:Q02AD}
\end{align}\label{eq:IQAT}
\end{subequations}

\noindent\emph{Proof} The starting point of our proof is similar to periodic case in Section \ref{sec:EQTS}. Using the definition of $Q_{1,1}$ in \eqref{eq:Q11AD}, the left hand side of \eqref{eq:Qsystem1stAD} reads
\begin{equation}
    \begin{aligned}
        &\beta \Qzz^{[+]} \left(t^{[2-]} Q_{1,0} - t^{-1} Q^{[2-]}_{1,0}\right) + \alpha \Qzz^{[-]} \left(t Q^{[2+]}_{1,0} - (t^{-1})^{[2+]} Q_{1,0}\right) - \alpha \beta c(t) \Qzz^{[+]} \Qzz^{[-]}.
    \end{aligned}
\end{equation}
From the identification rule \eqref{eq:IQ0Q1}, we have 
  \begin{align*}
 &\,\beta t^{[2-]} T_0^{[+]} Q_{1,0} - \alpha (t^{-1})^{[2+]} T_0^{[-]} Q_{1,0}- \beta t^{-1} T_0^{[+]} Q^{[2-]} + \alpha t T_0^{[-]} Q^{[2+]}  - \alpha \beta c(t) T_0^{[+]} T_0^{[-]}\\\nonumber
 =&\,(\beta t^{[2-]} T_0^{[+]}  - \alpha (t^{-1})^{[2+]} T_0^{[-]}) Q_{1,0}+q^{-\frac{1}{2}}T_{1/2}Q_{1,0}\\\nonumber
 =&\,(\beta t^{[2-]} T_0^{[+]}  - \alpha (t^{-1})^{[2+]} T_0^{[-]}+q^{-\frac{1}{2}}T_{1/2}) Q_{1,0}
    \end{align*}
where in the first equality we used the $TQ$-relation \eqref{eq:TQAD}. In the last line, one finds that the expression in the bracket is precisely $Q_{0,1}$ defined in \eqref{eq:Q01AD}, therefore we have proved the $QQ$-relation \eqref{eq:Qsystem1stAD}.

We can visualize the formula using the notation in Fig.~\ref{fig:Young_diagram}, where we focus on the first box at the left lower corner, 
\begin{align}
\raisebox{-.5\height}{
  \begin{tikzpicture}[box/.style={draw, rectangle, minimum size=1cm},
  dot/.style={circle, fill=black, inner sep=0pt, minimum size=3pt}]
  \node[box] (boxlhs1) at (0,0) {};
  \node[dot] at (boxlhs1.north west) {};
  \node at (0.6,0) {$\cdot$};
  \node[box] (boxlhs2) at (1.2,0) {};
  \node[dot] at (boxlhs2.south east) {};
  \node at (2.2,0) {$=$};
  \node at (2.6,0) {$\beta$};
  \node[box] (boxrhs11) at (3.4,0) {};
  \node[dot] at (boxrhs11.south west) {};
  \node at (3.2,-0.2) {$+$};
  \node at (4.0,0) {$\cdot$};
  \node[box] (boxrhs12) at (4.6,0) {};
  \node[dot] at (boxrhs12.north east) {};
  \node at (4.8,0.3) {$-$}; 
  \node at (5.6,0) {$+$};
  \node at (6,0) {$\alpha$};
  \node[box] (boxrhs21) at (6.8,0) {};
  \node[dot] at (boxrhs21.south west) {};
  \node at (6.6,-0.3) {$-$};
  \node at (7.4,0) {$\cdot$};
  \node[box] (boxrhs22) at (8,0) {};
  \node[dot] at (boxrhs22.north east) {};
  \node at (8.2,0.2) {$+$}; 
  \node at (9,0) {$-$};
  \node at (9.8,0) {$c(t)$};
  \node[box] (boxrhs31) at (10.8,0) {};
  \node[dot] at (boxrhs31.south west) {};
  \node at (10.6,-0.3) {$-$};
  \node at (11.4,0) {$\cdot$};
  \node[box] (boxrhs32) at (12,0) {};
  \node[dot] at (boxrhs32.south west) {};
  \node at (11.8,-0.2) {$+$};
  \end{tikzpicture}
  } \,.
\label{eq:Qsystemfirstbox}
\end{align}
Note that equation \eqref{eq:Qsystemfirstbox} differs from the periodic case \eqref{eq:QQPBC} by an additional inhomogeneous term that is proportional to $c(t)$, and the relative sign of the first two terms is also different. \par

The procedure for proving the $QQ$-relation \eqref{eq:Qsystem2ndAD}  is similar to the periodic case. We need first to derive a `$TQ$'-like relation for $T_1$, and then use the $TQ$-relation \eqref{eq:TQAD} and fusion relations \eqref{eq:fusionAD1}. The calculation is straightforward but tedious; thus we delegate it in Appendix~\ref{app:QQ_relation_2nd_box}. \par

It is straightforward to observe that the ZRC for both $Q_{0,1}$ and $Q_{0,2}$ from \eqref{eq:Qsystem1stAD} and \eqref{eq:Qsystem2ndAD} are equivalent to the ones of $T_{1/2}$ and $T_1$ from the $TQ$ and fusion relations. We conjecture that solving this $Q$-system of $Q_{0,n}$ and $Q_{1,n}$ with $n \in \{ 0,1,2\}$ is enough to obtain all physical solutions of the XXZ spin chain with anti-diagonal boundary conditions. We verify this by numerically solving the $Q$-systems and comparing to the results from exact diagonalization of the transfer matrix.

\subsection{Numerical results}
\label{subsec:ADnumeric}
In this subsection, we present numerical results for solving the $Q$-system~\eqref{eq:Qsystem1stAD} and \eqref{eq:Qsystem2ndAD}. In Table~\ref{table:twist poly}, we present the solution of the $Q$-system for the parameters $( L, \eta, \alpha, \beta)=\left(3,\log2,1,1\right)$. There are indeed $2^L=8$ physical solutions. Each solutions corresponds to a $Q$-polynomial of degree $L=3$. On the contrary, solving the $TQ$-relation leads to 12 solutions, which implies that 4 out of 12 solutions are unphysical.

We have a comment on the $\mathbb{Z}_2$ symmetry. For each solution $Q(t)$, we can uniquely find its corresponding $T(t)$. We observe that the $T$-polynomials fall into 4 groups, each group contains two $T$-polynomials, which only differ by a global sign. For example, we have
\begin{equation}
\begin{split}
&Q(t)=\frac{1}{8} \left( t^3-t^{-3}-4.4210 (t-t^{-1}) \right) \quad \leftrightarrow \quad  T(t)=0.6349( t^2+t^{-2})-1.1653\; , \\
&Q(t)=\frac{1}{8} \left( t^3+t^{-3}-6.5790 (t-t^{-1}) \right) \quad \leftrightarrow \quad T(t)=    -0.6349 (t^2+t^{-2})+1.1653\,.
\end{split}
\end{equation} 
As we can see, the zeroes for two $Q$-polynomials are different for the $T$-polynomials that differ only by a global sign. This is because the Hamiltonian \eqref{eq:HZ2} with $\alpha = \beta=1$ is $\mathbb{Z}_2$-invariant but the transfer matrix is not, as previously mentioned. 

\begin{table}[H]
    \centering
    \begin{tabular}{|c|c|}
    \hline
    & $Q$ polynomials  \quad $2^{L} Q(t)$  \\
\hline
1 & $t^3 -4.4210 t + 4.4210 t^{-1} - 1.0000 t^{-3}$ \\
\hline

2 & $t^3 -10.829 t + 10.829 t^{-1} - 1.0000 t^{-3}$ \\
\hline

3 & $t^3 -6.5790 t + 6.5790 t^{-1} - 1.0000 t^{-3}$\\
\hline

4 & $t^3 -0.17100 t + 0.17100 t^{-1} - 1.0000 t^{-3}$ \\
\hline

5 & $t^3 +(7.9464-2.0104 \mathrm{i}) t + (3.1250-7.5777 \mathrm{i}) t^{-1} - (0.14286-0.98974 \mathrm{i}) t^{-3}$ \\
\hline

6 & $t^3 -(7.9464+2.0104 \mathrm{i}) t + (3.1250+7.5777 \mathrm{i}) t^{-1} - (0.14286+0.98974 \mathrm{i}) t^{-3}$\\
\hline

7 & $t^3 -(4.6483-1.9183 \mathrm{i}) t + (3.0436-4.0028 \mathrm{i}) t^{-1} - (0.25581-0.96673 \mathrm{i}) t^{-3}$ \\
\hline

8 & $t^3 -(4.6483+1.9183 \mathrm{i}) t + (3.0436+4.0028 \mathrm{i}) t^{-1} - (0.25581+0.96673 \mathrm{i}) t^{-3}$ \\
\hline

\textcolor{gray}{9}
& {\color{gray}$t^3 -3.5000 t+ 3.5000 t^{-1} -1.0000 t^{-3}$} \\
\hline

\textcolor{gray}{10}
& {\color{gray}$t^3 -1.5000 t- 1.5000 t^{-1} -1.0000 t^{-3}$} \\
\hline

\textcolor{gray}{11}
& {\color{gray}$t^3 -2.8008 t+ 1.7521 t^{-1} -0.30083 t^{-3}$}\\
\hline

\textcolor{gray}{12}
& {\color{gray}$t^3 -5.8242 t+ 9.3104 t^{-1} -3.3242 t^{-3}$}\\
\hline
    \end{tabular}
    \caption{$Q$-polynomials for the parameters $( L, \eta, \alpha, \beta)=\left(3,\log2,1,1\right)$. Solutions 1--8 are obtained from solving the $Q$-system~\eqref{eq:Qsystem1stAD} and \eqref{eq:Qsystem2ndAD}, or equivalently the `$TQ$+fusion' system~\eqref{eq:TQAD} and \eqref{eq:fusionAD1}, all of which are physical. Solutions 9--12 are additional solutions to the $TQ$-relation~\eqref{eq:TQAD}, which are unphysical.}\label{table:twist poly}
\end{table}

\section{XXZ chain with non-diagonal boundary magnetic field}
\label{sec:unparallel}

In this section, we establish the $Q$-system for XXZ spin chain with non-diagonal boundary magnetic fields, which is another typical situation without $U(1)$ symmetry. We follow the same strategy as the previous section and derive the minimal $Q$-system from $TQ$- and fusion relations.

\subsection{Review of diagonal case}

Before discussing the $Q$-systems without $U(1)$ symmetry, we first review of $Q$-systems with diagonal boundary magnetic fields to highlight the differences between the open and closed cases.

The Hamiltonian of XXZ spin-$\frac{1}{2}$ chain of length $L$ with diagonal boundary magnetic fields is~\cite{Nepomechie:2019gqt} 
\begin{equation}
\label{eq:boundaryParallel}
\mathbf{H}=\sum_{j=1}^{L-1} \left[ \sigma^x_j \sigma^x_{j+1} + \sigma^y_j \sigma^y_{j+1} + \Delta \sigma^z_j \sigma^z_{j+1} \right]-\sinh(\eta)\coth(\beta\eta)\sigma_1^z+\sinh(\eta)\coth(\alpha
    \eta)\sigma_L^z,
\end{equation}
where $\alpha$ and $\beta$ are arbitrary parameters. This model is clearly $U(1)$ invariant. 

The $R$-matrix is defined as \eqref{eq:R_mat_XXZ} and the $K$-matrices which describe the boundary magnetic fields in the algebraic Bethe ansatz are given by 
\begin{align}
\mathbf{K}^L(u)=&\,\begin{pmatrix}
        \sinh(u+\alpha\eta) & 0 \\
        0 & \sinh(-u+\alpha\eta)
    \end{pmatrix},\\\nonumber
\mathbf{K}^R(u)=&\,\begin{pmatrix}
        \sinh(-u+\beta\eta-\eta) & 0 \\
        0 & \sinh(u+\beta\eta+\eta)
    \end{pmatrix}\,.
\end{align}
These $K$-matrices are diagonal and lead to the $U(1)$ preserving Hamiltonian \eqref{eq:boundaryParallel}. The corresponding monodromy matrix $\mathbf{M}_a(u)$ and the transfer matrix $\mathbf{T}(u)$ are defined as
\begin{equation}
   \begin{aligned}
        &\mathbf{M}_a(u,\eta,\{\alpha,\beta\})=\mathbf{K}^L_a(u)\prod^L_{n=1}\mathbf{R}_{an}(u)\mathbf{K}^R_a(u)\prod^1_{m=L}\mathbf{R}_{am}(u)\;,\\
        &\mathbf{T}(u,\eta,\{\alpha,\beta\})=\text{Tr}_a\left[\mathbf{M}_a(u-\tfrac{\eta}{2},\eta,\{\alpha,\beta\})\right]\;.
   \end{aligned}
\end{equation}
Following the standard procedure, we can use algebraic Bethe ansatz to diagonalize the transfer matrix. The Bethe ansatz equations for the Bethe roots read
\begin{equation}
    \begin{aligned}
        \frac{g^{[-]}(u_j)}{f^{[+]}(u_j)} \left(\frac{\sinh(u_j+\frac{\eta}{2})}{\sinh(u_j-\frac{\eta}{2})}\right)^{2L}=\prod^M_{k=1, k\neq j}\frac{\sinh(u_j-u_k+\eta)\sinh(u_j+u_k+\eta)}{\sinh(u_j-u_k-\eta)\sinh(u_j+u_k-\eta)} \;, \\ j\in \{1,2,...,M\},\quad M\in\{0,...,L\} \;,
    \end{aligned}
\end{equation}
where $M$ is the magnon number and the two functions $f(u)$ and $g(u)$ are defined as
\begin{equation}
    \begin{aligned}
        &f(u)=\sinh(u-\alpha\eta)\sinh(u+\beta\eta) \; ,\\
        &g(u)=f(-u)=\sinh(u+\alpha\eta)\sinh(u-\beta\eta) \; ,
    \end{aligned}
\end{equation}
which encode the information of the boundary magnetic fields.

The $TQ$-relation in this case becomes
\begin{align}
\label{eq:TQhomoOpen}
    -\sinh(2u)T(u)Q(u)=&\,\sinh(2u+\eta)T_0^{[+]}(u)g^{[-]}(u)Q^{[2-]}(u)\\\nonumber
    &\,+\sinh(2u-\eta)T_0^{[-]}(u)f^{[+]}(u)Q^{[2+]}(u),
\end{align}
with
\begin{equation}
    T_0 (u)= \sinh^{2L} (u), \qquad  Q(u)=\prod_{j=1}^L\sinh(u-u_j)\sinh(u+u_j).
\end{equation}
Since the model preserves $U(1)$ symmetry, the $TQ$-relation \eqref{eq:TQhomoOpen} is homogeneous.\par

The corresponding $Q$-systems of the model has been established in~\cite{Nepomechie:2019gqt}. The $QQ$-relations read
\begin{equation}
    \begin{aligned}
        &\sinh(2u)Q_{1,n}= f^{[-(n-1)]}Q^+_{1,n-1}-g^{[n-1]}Q^-_{1,n-1},\\
        &\sinh(2u)Q_{0,n}Q_{1,n-1}= Q^+_{1,n}Q^-_{0,n-1}-Q^-_{1,n}Q^+_{0,n-1},\quad n=1,2,....
    \end{aligned}
\end{equation}
where $Q_{0,0}(u)=T_0(u)$ and $Q_{1,0}(u)=Q(u)$.
The $Q$-system reproduce the complete spectrum of the transfer matrix, which has been checked numerically.

\subsection{\texorpdfstring{$Q$-system with non-diagonal boundary magnetic field}{Q-systems with non-diagonal boundary magnetic field}}
Now we consider XXZ spin chain with non-diagonal boundary magnetic fields.\footnote{The XXX spin chain with non-diagonal boundary magnetic fields corresponds to the isotropic limit $\eta\to 0$ and will be discussed separately in Appendix~\ref{app:XXXnondiagonal}.}
The transfer matrix for open XXZ spin chain is constructed by standard procedure~\cite{EKSklyanin_1988} with the 6-vertex $R$-matrix \eqref{eq:R_mat_XXZ} and the following boundary $K$-matrix~\cite{Nepomechie_2003,GHOSHAL_1994,Vega_1993}
\begin{equation}
    \mathbf{K}_a (u,\alpha, \beta, \theta) = \begin{pmatrix}
        K_{11} (u,\alpha, \beta) & K_{12} (u, \theta) \\
        K_{21} (u, \theta) & K_{22} (u,\alpha, \beta) 
    \end{pmatrix} \; ,
\end{equation}
with the matrix elements
\begin{equation}
\begin{split}
    & K_{11} (u,\alpha, \beta ) = 2 \left[ \sinh \alpha \cosh \beta \cosh u + \cosh \alpha \sinh \beta \sinh u \right] \; , \\
    & K_{22} (u,\alpha, \beta ) = 2 \left[ \sinh \alpha \cosh \beta \cosh u - \cosh \alpha \sinh \beta \sinh u \right] \; , \\
    & K_{12} (u,  \theta) = e^{\theta} \sinh (2u) \; , \quad K_{21} (u,  \theta) = e^{-\theta} \sinh (2u) \; .
\end{split}
\end{equation}
The $K$-matrices at the two boundaries are\footnote{Note the additional minus sign in front of parameters $\alpha_+$ and $\beta_+$.}
\begin{equation}
    \mathbf{K}_a^{-} (u) : = \mathbf{K}_a (u,\alpha_- , \beta_- , \theta_- ) \; , \quad \mathbf{K}_a^{+} (u) : = \mathbf{K}_a (- u - \eta ,- \alpha_+ , -\beta_+ , \theta_+ ) \; .
\end{equation}
where $\eta$  is the bulk anisotropy parameter and $\alpha_\pm,\beta_\pm,\theta_\pm$ are boundary parameters. The monodromy matrix is defined as
\begin{equation}
\begin{split}
    \mathbf{M}_a (u,\eta , \{ \alpha_\pm , \beta_\pm , \theta_\pm \} )  = \mathbf{K}_a^{+} (u) \prod_{n=L}^1 \mathbf{R}_{an} (u,\eta) \mathbf{K}_a^{-} (u) \prod_{m=1}^L \mathbf{R}_{m a} (u,\eta) \; ,
\end{split}
\end{equation}
and the transfer matrix is defined as
\begin{equation}
    \mathbf{T} (u,\eta , \{ \alpha_\pm , \beta_\pm , \theta_\pm \} ) = \mathrm{Tr}_a \left[ \mathbf{M}_a \left( u - \frac{\eta}{2} ,\eta , \{ \alpha_\pm , \beta_\pm , \theta_\pm \} \right)  \right] \;,
\end{equation}
from which we can derive the Hamiltonian of the spin chain~\cite{Nepomechie_2003,Wang:2015off}
\begin{equation}
    \mathbf{H} = \sum_{j=1}^{L-1} \left[ \sigma^x_j \sigma^x_{j+1} + \sigma^y_j \sigma^y_{j+1} + \Delta \sigma^z_j \sigma^z_{j+1} \right] + \vec{h}_1 \cdot \vec{\sigma}_1 + \vec{h}_L \cdot \vec{\sigma}_L \; ,
\end{equation}
with boundary magnetic fields given by
\begin{equation}
\begin{split}
    \vec{h}_1 = \frac{\sinh \eta}{\sinh \alpha_- \cosh \beta_- } \left( \cosh \theta_- , \mathrm{i} \sinh \theta_- , \cosh \alpha_- \sinh \beta_- \right) \; , \\ 
    \vec{h}_L = \frac{\sinh \eta}{\sinh \alpha_+ \cosh \beta_+ } \left( \cosh \theta_+ , \mathrm{i} \sinh \theta_+ , - \cosh \alpha_+ \sinh \beta_+ \right) \; .
\end{split}
\end{equation}

The Bethe roots $\{u_j\}_{j=1}^{2L}$, which label eigenstates of the transfer matrix, satisfy the following BAE
\begin{equation}
    \frac{f\big(u_j-\frac{\eta}{2}\big)Q(u_j-\eta)}{\sinh (2u_j-\eta) \sinh^{2L} \big(u_j-\frac{\eta}{2}\big)} + \frac{g\big(u_j+\frac{\eta}{2}\big)Q(u_j+\eta)}{\sinh (2u_j+\eta) \sinh^{2L} \big(u_j+\frac{\eta}{2}\big)} = \frac{x}{2} \; ,
\label{eq:BAEnondiagonal}
\end{equation}
where two boundary polynomials are
\begin{equation}
    f (u) = \sinh(u-\alpha_+) \sinh(u-\alpha_-) \cosh (u-\beta_+) \cosh (u-\beta_-) = g(-u) \; ,
\label{eq:boundarypolys}
\end{equation}
and the $Q$-polynomial is
\begin{equation}
    Q(u) = \prod_{j=1}^L \sinh (u-u_j) \sinh (u+u_j) \; .
\label{eq:Qpolyopen}
\end{equation}
There is an additional constant term on the right-hand side, which reads
\begin{equation}
    x = \cosh[ \alpha_+ + \beta_+ + \alpha_- + \beta_- + (L+1) \eta ] - \cosh (\theta_- - \theta_+ ) \; .
\label{eq:Nepomechiecond}
\end{equation}

We would like to remark that $x=0$ is known as the \emph{Nepomechie condition}~\cite{Nepomechie:2003vv} in the literature. When the condition is satisfied, the BAE become the same as the BAE of the spin chain with parallel magnetic field. The number of Bethe roots for each solution is not $2L$ in that case, cf. \cite{Chernyak:2022ejj}. In this paper, we focus on the generic scenario, i.e. the Nepomechie condition is \emph{not} satisfied ($x\neq 0$). Hence, we have $(2L)$-many Bethe roots in \eqref{eq:Qpolyopen} in this section.

\paragraph{The `$TQ$+fusion' system}

The inhomogeneous $TQ$-relation has been obtained in~\cite{Nepomechie_2003} and is given by
\begin{equation}\label{eq:open_XXZ_TQ}
\begin{split}
     \sinh(2u) T Q = &- 4 \left[ \sinh(2u+\eta) f^{[-]} T_0^{[+]} Q^{[2-]} + \sinh(2u-\eta) g^{[+]} T_0^{[-]} Q^{[2+]}  \right] \\
    & + 2 x T_0^{[+]} T_0^{[-]} \sinh(2u+\eta) \sinh (2u) \sinh(2u-\eta) \; ,
\end{split}
\end{equation}
where
\begin{align}
    & T_0 (u) = \sinh^{2L}(u)\;, \qquad  Q(u)=\prod_{j=1}^L\sinh(u-u_j)\sinh(u+u_j)\;,\label{eq:def_open_Q_func}
\end{align}
and the boundary polynomials are given in \eqref{eq:boundarypolys}. 
The fusion relation is given by~\cite{Nepomechie_2003,Miao_2021,mezincescu1992fusion}
\begin{equation}
\label{eq:open_XXZ_fusion}
\begin{split}
    (\sinh(2u) T_0) (\sinh(2u) T_1) = & (-\sinh(2u) T_{1/2})^{[+]} (-\sinh(2u) T_{1/2})^{[-]} \\
    & -16 f g (\sinh(2u) T_0)^{[2+]} (\sinh(2u) T_0)^{[2-]} \;.
\end{split}
\end{equation}

\paragraph{Minimal $Q$-system}
The minimal $Q$-system for the model is given by the following $QQ$-relations
\begin{align}
 &\sinh(2u)Q_{1,1} = g Q^{[+]}_{1,0} - f Q^{[-]}_{1,0}\\
 &\sinh(2u)Q_{1,2} = g^{[-]} Q^{[+]}_{1,1} - f^{[+]} Q^{[-]}_{1,1}
\end{align}
\begin{align}
&\label{eq:QQ00Open}\sinh(2u)\Qzo\Qoz=\Qzz^{[+]} \Qoo^{[-]} - \Qzz^{[-]} \Qoo^{[+]}+\frac{1}{2}x\sinh(2u)\Qzz^{[+]}\Qzz^{[-]},\\
&\sinh(2u)\Qzt\Qoo=\Qzo^{[+]} \Qot^{[-]} - \Qzo^{[-]} \Qot^{[+]} +\frac{1}{2}x\left(gQ_{0,0}^{[2+]}Q_{0,1}^{[-]}-fQ_{0,0}^{[2-]}Q_{0,1}^{[+]}\right).\label{eq:QQ01Open}
\end{align}
It can be derived from the $TQ$+fusion system via the following identification 
\begin{align}
Q_{0,0}=&\,T_0\;,\qquad Q_{1,0}=Q\;,\\\nonumber
\Qzo=&\,\frac{1}{4U^{[+]}U^{[-]}}T_{1/2}+\frac{1}{U^{[-]}U}g^{[-]}T_0^{[+]}+\frac{1}{UU^{[+]}}f^{[+]}T_0^{[-]}\;,\label{eq:def_open_Qzo}\\
\Qzt=&\,\frac{1}{16U^{[2-]}U^{[-]}U^{[+]}U^{[2+]}}T_1\\\nonumber
&\,+\frac{U^{[2-]}+U}{4U^{[2-]}U^{[-]}U^2U^{[2+]}}g^{[2-]}T_\frac{1}{2}^{[+]}+\frac{U+U^{[2+]}}{4U^{[2-]}U^2U^{[+]}U^{[2+]}}f^{[2+]}T_\frac{1}{2}^{[-]}\\\nonumber 
&\,+\frac{gT_0^{[2+]}}{U^{[-]}U^2U^{[+]}}+\frac{fT_0^{[2-]}}{U^{[-]}U^2U^{[+]}}+\frac{U^{[2-]}f^{[2+]}+U^{[2+]}g^{[2-]}+Uf^{[2+]}g^{[2-]}}{U^{[2-]}U^{[-]}UU^{[+]}U^{[2+]}}T_0\; ,
\end{align}
where we use $U = \sinh(2u)$.\par

\noindent\emph{Proof} We first define the $Q_{1,n}$-functions using the recursion relation
\begin{align}
\sinh(2u)Q_{1,n}=g^{[(n-1)-]}Q_{1,n-1}^{[+]}-f^{\left[(n-1)+\right]}Q_{1,n-1}^{[-]}\,.
\end{align}
In particular,
\begin{align}
  &UQ_{1,1} = g Q^{[+]}_{1,0} - f Q^{[-]}_{1,0} \label{eq:def_open_Q11}\,,\\
  &UQ_{1,2} = g^{[-]} Q^{[+]}_{1,1} - f^{[+]} Q^{[-]}_{1,1}\,,\label{eq:def_open_Q12}  
\end{align}
where we have introduced the shorthand notation $U(u)=\sinh(2u)$. By applying $TQ$-relation \eqref{eq:open_XXZ_TQ}, we find
\begin{equation}
    \begin{aligned}
    &\Qzz^{[+]} \Qoo^{[-]} + \Qzz^{[-]} \Qoo^{[+]} \\&=  \Qzz^{[+]}\frac{1}{U^{[-]}}\left(g^{[-]}\Qoz-f^{[-]}\Qoz^{[2-]}\right)-\Qzz^{[-]}\frac{1}{U^{[+]}}\left(g^{[+]}\Qoz^{[2+]}-f^{[+]}\Qoz\right) \\
    & =U\left(\frac{1}{4U^{[+]}U^{[-]}}T_{1/2}+\frac{1}{U^{[-]}U}g^{[-]}T_0^{[+]}+\frac{1}{UU^{[+]}}f^{[+]}T_0^{[-]}\right)\Qoz-\frac{1}{2}xU\Qzz^{[+]}\Qzz^{[-]}.
    \end{aligned}
\end{equation}
This leads to a $QQ$-relation with an additional `source term'
\begin{equation}
    U\Qzo\Qoz=\Qzz^{[+]} \Qoo^{[-]} - \Qzz^{[-]} \Qoo^{[+]}+\frac{1}{2}xU\Qzz^{[+]}\Qzz^{[-]}.
\end{equation}
This is nothing but \eqref{eq:QQ00Open}. To proceed, we multiply both sides of the fusion relation \eqref{eq:open_XXZ_fusion} with $\Qoo$
\begin{equation}\label{eq:UT0UT1Q11}
    \begin{aligned}
        (U T_0) (U T_1)\Qoo= & (-U T_{1/2})^{[+]} (-U T_{1/2})^{[-]}\frac{1}{U}\left(g Q^{[+]}_{1,0} - f Q^{[-]}_{1,0}\right)\\& -16 f g (U T_0)^{[2+]} (U T_0)^{[2-]}\frac{1}{U}\left(g Q^{[+]}_{1,0} - f Q^{[-]}_{1,0}\right) .
    \end{aligned}
\end{equation}
Applying the $TQ$-relation repeatedly to the right-hand side of \eqref{eq:UT0UT1Q11}, one finds that the result is proportional to $T_0$. Dividing both sides of \eqref{eq:UT0UT1Q11} by $T_0$ gives~\footnote{More details can be found in Appendix~\ref{app:derivation_open_case}.}
\begin{equation}\label{eq:open_T1Q11}
    \begin{aligned}
       U^2T_1\Qoo=&16fg\left[f^{[2-]}U^{[2+]}T_0^{[2+]}\Qoz^{[3-]}-g^{[2+]}U^{[2-]}T_0^{[2-]}\Qoz^{[3+]}\right]\\&+4\left[ff^{[2-]}U^{[+]}T_\frac{1}{2}^{[+]}\Qoz^{[3-]}-gg^{[2+]}U^{[-]}T_\frac{1}{2}^{[-]}\Qoz^{[3+]}\right]+S(x),
    \end{aligned}
\end{equation}
where $S(x)$ is
\begin{equation}
    \begin{aligned}
        S(x)&=8xfgU^{[2-]}U^{[2+]}\left(U^{[+]}-U^{[-]}\right)T_0^{[2-]}T_0^{[2+]}\\&+2xU^{[-]}U^{[+]}\left[gU^{[2+]}T_\frac{1}{2}^{[-]}T_0^{[2+]}-fU^{[2-]}T_\frac{1}{2}^{[+]}T_0^{[2-]}\right]\,.
    \end{aligned}
\end{equation}
Now we derive the rest $QQ$-relation. By using \eqref{eq:def_open_Q12}, we have
\begin{equation}\label{eq:Q01Q12}
    \begin{aligned}
       \Qzo^{[+]} \Qot^{[-]} - & \Qzo^{[-]} \Qot^{[+]} =\frac{\Qzo^{[+]}}{U^{[-]}}\left[g^{[2-]}\Qoo-f\Qoo^{[2-]}\right]-\frac{\Qzo^{[-]}}{U^{[+]}}\left[g\Qoo^{[2+]}-f^{[2+]}\Qoo\right]\\&=-\left[\frac{g\Qzo^{[-]}\Qoo^{[2+]}}{U^{[+]}}+\frac{f\Qzo^{[+]}\Qoo^{[2-]}}{U^{[-]}}\right]+\left[\frac{g^{[2-]}\Qzo^{[+]}}{U^{[-]}}+\frac{f^{[2+]}\Qzo^{[-]}}{U^{[+]}}\right]\Qoo.
    \end{aligned}
\end{equation}
Using \eqref{eq:def_open_Qzo} and \eqref{eq:left_bracket} from Appendix~\ref{app:derivation_open_case}, we can simplify the right hand side of the above equation and obtain
\begin{equation}\label{eq:Q01Q12_final}
    \begin{aligned}
        \Qzo^{[+]} \Qot^{[-]} - \Qzo^{[-]} \Qot^{[+]} &=U\Qzt\Qoo+V(x)\;,
    \end{aligned}
\end{equation}
where the inhomogeneous term reads
\begin{equation}
    \begin{aligned}
        V(x)&=-\frac{S(x)}{16U^{[2-]}U^{[-]}UU^{[+]}U^{[2+]}}-\frac{xgg^{[2-]}T_0T_0^{[2+]}}{2U^{[2-]}U^{[-]}}+\frac{xff^{[2+]}T_0T_0^{[2-]}}{2U^{[+]}U^{[2+]}}\\&=-\frac{1}{2}x\left(gQ_{0,0}^{[2+]}Q_{0,1}^{[-]}-fQ_{0,0}^{[2-]}Q_{0,1}^{[+]}\right)\;.
    \end{aligned}
\end{equation}
Therefore 
\begin{align}
\Qzo^{[+]} \Qot^{[-]} - \Qzo^{[-]} \Qot^{[+]} &=U\Qzt\Qoo-\frac{1}{2}x\left(gQ_{0,0}^{[2+]}Q_{0,1}^{[-]}-fQ_{0,0}^{[2-]}Q_{0,1}^{[+]}\right)\;,
\end{align}
which is equivalent to \eqref{eq:QQ01Open}. This completes the proof.

Following similar steps as above, it is straightforward although a bit tedious to express higher $Q$-functions in terms of higher $T$-functions and obtain more $QQ$-relations.\par

In terms of multiplicative variables $t=e^u$ and $q=e^{\eta}$, the $QQ$-relations become
\begin{align}
    &\frac{t^2-t^{-2}}{2}Q_{1,1} = g Q^{[+]}_{1,0} - f Q^{[-]}_{1,0} \; , \\
    &\frac{t^2-t^{-2}}{2}Q_{1,2} = g^{[-]} Q^{[+]}_{1,1} - f^{[+]} Q^{[-]}_{1,1} \,.
\end{align}
\begin{align}
    &\frac{t^2-t^{-2}}{2}\Qzo\Qoz=\Qzz^{[+]} \Qoo^{[-]} - \Qzz^{[-]} \Qoo^{[+]}+\frac{t^2-t^{-2}}{4}x\Qzz^{[+]}\Qzz^{[-]} \; ,\\
    &\frac{t^2-t^{-2}}{2}\Qzt\Qoo=\Qzo^{[+]} \Qot^{[-]} - \Qzo^{[-]} \Qot^{[+]} +\frac{1}{2}x\left(gQ_{0,0}^{[2+]}Q_{0,1}^{[-]}-fQ_{0,0}^{[2-]}Q_{0,1}^{[+]}\right) \; .
\end{align}

In terms of variable $t$, we have
\begin{equation}
    Q_{0,0}(t)=\frac{1}{2^{2L}}\left(t-\frac{1}{t}\right)^{2L} \; ,
\end{equation}
\begin{equation}
\begin{split}
    f(t)& =\frac{\left(te^{-\alpha_+}-t^{-1}e^{\alpha_+}\right)\left(te^{-\alpha_-}-t^{-1}e^{\alpha_-}\right)\left(te^{-\beta_+}-t^{-1}e^{\beta_+}\right)\left(te^{-\beta_-}-t^{-1}e^{\beta_-}\right)}{16} \\ 
    & = g\left(t^{-1}\right) \; .
\end{split}
\end{equation}

\subsection{Numerical Results}
To solve the minimal $Q$-system, we require that all the $Q$-functions are Laurent polynomials in variable $t$. Specifically, we make an ansatz for $Q_{1,0}$
\begin{equation}
    Q_{1,0}(t)=\frac{1}{2^{2L}}\left(\sum_{k=0}^{L-1}c_k(t^{2k}+t^{-2k})+t^{2L}+t^{-2L}\right),
\end{equation}
where $c_k$ are coefficients to be determined. From the $QQ$-relation, we can obtain the expressions of $\Qzo$ and $\Qzt$
\begin{equation}
\begin{split}
    &\Qzo=\frac{2\Qzz^{[+]} \Qoo^{[-]} - 2\Qzz^{[-]} \Qoo^{[+]}+\frac{1}{2}x\left(t^2-t^{-2}\right)\Qzz^{[+]}\Qzz^{[-]}}{\left(t^2-t^{-2}\right)\Qoz} \; ,\\
    &\Qzt=\frac{2\Qzo^{[+]} \Qot^{[-]} - 2\Qzo^{[-]} \Qot^{[+]} +x\left(gQ_{0,0}^{[2+]}Q_{0,1}^{[-]}-fQ_{0,0}^{[2-]}Q_{0,1}^{[+]}\right)}{\left(t^2-t^{-2}\right)\Qoo} \; .
\end{split}
\label{eq:quotientofQQ}
\end{equation}
The right-hand sides of the two equations might not be Laurent polynomials in $t$. Hence, we further require the remainders of the right-hand sides to vanish, leading to the ZRC. Then we solve the ZRC, obtaining the polynomial $Q_{1,0}(t)$. The zeros of $Q_{1,0}(t)$ after taking the logarithm are the Bethe roots. We numerically solved the $Q$-system for various choices of parameters and verified that they give all and only physical solutions.

For the most generic case~\footnote{\emph{i.e.} $q=e^\eta$ is not at root of unity, and ``Nepomechie condition'' $x=0$ of \eqref{eq:Nepomechiecond} is not satisfied.} that all the boundary parameters are non-zero. As an example, we choose the parameters to be
\begin{align} 
( L, \eta, \alpha_ +, \alpha_-, \beta_+, \beta_-, \theta_+, \theta_-)=\left(3,\log2,1,2,\frac{1}{3},\frac{1}{4},\frac{1}{3},\frac{1}{2}\right)\,.
\end{align}
The numerical solutions of the $Q$-system are presented in Table~\ref{table:open_case}. 
Since the total Hilbert space is $\mathcal{H}_L =(\mathbb{C}^2)^{\otimes L}$, the solutions of the $Q$-system should be 8 distinct $Q$-polynomials, where each $Q$-functions \eqref{eq:def_open_Q_func} has 12 different zeros.

\begin{table}[H]
 \centering
\begin{tabular}{|c|c|}
\hline
& $Q$ polynomials  \quad $2^{2L} Q(t)$  \\
\hline
1 & $t^6+t^{-6}+69.75177\left(t^2+t^{-2}\right) - 14.03306\left(t^4+t^{-4}\right)+24.75197$   \\
\hline
2 & $t^6+t^{-6}+90.32738\left(t^2+t^{-2}\right)  - 14.34073\left(t^4+t^{-4}\right)-336.7673$      \\
\hline
3 & $t^6+t^{-6}+55.21936\left(t^2+t^{-2}\right) - 13.51494\left(t^4+t^{-4}\right)-48.85021$      \\
\hline
4 & $t^6+t^{-6}+67.92860 \left(t^2+t^{-2}\right) - 13.95974\left(t^4+t^{-4}\right)-106.6805$      \\
\hline
5 &$t^6+t^{-6}+51.59731\left(t^2+t^{-2}\right) - 13.56352 \left(t^4+t^{-4}\right)-79.97889$  \\
\hline
6 &  $t^6+t^{-6}+79.55786\left(t^2+t^{-2}\right) - 13.72927\left(t^4+t^{-4}\right)-217.0126$ \\ 
\hline
7 &  $t^6+t^{-6}+61.79089\left(t^2+t^{-2}\right) - 12.50471\left(t^4+t^{-4}\right)-134.9551$\\
\hline
8 &  $t^6+t^{-6}+66.34321 \left(t^2+t^{-2}\right) - 12.88443\left(t^4+t^{-4}\right)-152.0944$ \\
\hline
\textcolor{gray}{9} &  {\color{gray} $t^6+t^{-6}+52.11624 \left(t^2+t^{-2}\right) - 13.57325\left(t^4+t^{-4}\right)-80.72930$ }\\
\hline
\textcolor{gray}{10} &  {\color{gray} $t^6+t^{-6}+26.56158  \left(t^2+t^{-2}\right) - 8.462315\left(t^4+t^{-4}\right)-38.56410$} \\
\hline
\textcolor{gray}{11} &  {\color{gray} $t^6+t^{-6}+(21.1595 - 5.0148  \mathrm{i})  \left(t^2+t^{-2}\right) - (7.3819 - 1.0029 \mathrm{i})\left(t^4+t^{-4}\right)- 29.6508 + 8.2745\mathrm{i}$} \\
\hline
\textcolor{gray}{12} &  {\color{gray} $t^6+t^{-6}+(21.1595 + 5.0148  \mathrm{i})  \left(t^2+t^{-2}\right) - (7.3819 + 1.0029 \mathrm{i})\left(t^4+t^{-4}\right)-29.6508 - 8.2745\mathrm{i}$} \\
\hline
\end{tabular}
\caption{$Q$ polynomials for the following parameters $( L, \eta, \alpha_ +, \alpha_-, \beta_+, \beta_-, \theta_+, \theta_-)=\left(3,\log2,1,2,\frac{1}{3},\frac{1}{4},\frac{1}{3},\frac{1}{2}\right)$. Solutions 1--8 are obtained from solving the $Q$-system~\eqref{eq:QQ00Open} and \eqref{eq:QQ01Open}, or equivalently the `$TQ$+fusion' system~\eqref{eq:open_XXZ_TQ} and \eqref{eq:open_XXZ_fusion}, all of which are physical. Solutions 9--12 are additional solutions to the $TQ$ relation~\eqref{eq:open_XXZ_TQ}, which are all unphysical.}
\label{table:open_case}
\end{table}

The numerical results show that the solutions obtained from the $Q$-system are entirely contained within the set of solutions from the $TQ$-relation. It demonstrates that the $Q$-system of the XXZ model with non-diagonal boundary magnetic fields yields all the physical solutions.

\section{Conclusions and discussions}
\label{sec:concl}

In this paper, we construct the rational $Q$-system for $U(1)$-symmetry breaking models, specifically the XXZ spin chain with anti-diagonal twist and non-diagonal boundary conditions. Our approach stems from a novel perspective on the  $Q$-system itself, demonstrating that it can be derived by combining the $TQ$-relation (both homogeneous and inhomogeneous forms) with the fusion relation. By identifying higher-spin $T$-functions with specific combinations of higher $Q$-functions, we successfully obtained the corresponding inhomogeneous $QQ$-relations. The proposed $Q$-system yields all, and only, physical solutions, a result we have verified numerically.

Viewing the $Q$-system as emerging from the interplay of the $TQ$-relation and the $T$-system provides a powerful guiding principle, applicable to a wide range of models for which both relations are known. A central, and not yet fully understood, question is why higher $T$-functions encode the physicality of a $TQ$-relation solution. For rational spin chains with periodic boundary conditions, this can be understood through the completeness of Wronskian Bethe equations. In that context, the fusion relations, formulated as Hirota dynamics, are solved by a finite set of $Q$-functions. These $Q$-functions satisfy a Wronskian-type relation. Under the assumption that all $Q$-functions are polynomials, it has been proven that these Wronskian Bethe equations provide the complete set of solutions to the Bethe Ansatz equations \cite{2007arXiv0706.0688M,Chernyak:2020lgw}. Since the polynomiality of all $Q$-functions implies the polynomiality of all $T$-functions, it is natural to posit that the polynomiality of $T$-functions serves as the physicality condition for the $TQ$-relation. For the more general models considered here, the derivation and the proof of completeness of the Wronskian Bethe equations remain open, and thus the connection between $T$-function polynomiality and BAE physicality stands as a conjecture. It would be highly interesting and important to prove (or correct) this conjecture more rigorously.

Meanwhile, testing this conjecture in broader contexts represents an important direction for future work. A natural next step involves higher spin models that break $U(1)$ symmetry. Preliminary studies suggest that in these cases, requiring only the first higher $T$-function, namely $T_1$ to be a polynomial is insufficient; one must also consider higher spin-dependent $T$-functions. A useful strategy would be to first re-derive the higher spin $Q$-system \cite{Hou:2023ndn} from our new perspective and then extend it to $U(1)$-symmetry breaking scenarios.

Further challenging and interesting generalizations lie ahead. Models with different underlying symmetries, such as those of $D_n$- and $E_n$-type, present a distinct case. The $QQ$-relations have been studied in \cite{Ferrando:2020vzk,Ekhammar:2020enr} and are known to differ from the $A_n$ case, the Wronskian Bethe equations have been proposed. Given that the $TQ$-relations and fusion relations for these models are known, it would be interesting to investigate whether combining them leads to a $Q$-system which yields complete solution of BAE.

Finally, we note the longstanding challenge of solving the BAE of the XYZ model~\cite{Baxter1973-ii, Baxter1973-iii}, for which an analog of the $Q$-system approach has not been established. Developing such a framework, guided by the principles outlined in this work, would be a significant and interesting goal.

\acknowledgments
We would like to thank the Innovation Academy for Precision Measurement Science and Technology, CAS for hospitality where the final stage of the work is done and part of the results are presented at the ``Challenges in Integrability'' workshop. The work of Y.J. is supported by Startup Fund No.4007022326 of Southeast University and National Natural Science Foundation of China through Grant No.12575073 and 12247103. The work of Y.M. is supported by the World Premier International Research Center Initiative (WPI), MEXT, Japan and the UTokyo Global Activity Support Program for Young Researchers.

\newpage

\appendix

\section{Polynomiality of higher-spin transfer matrices}
\label{app:recursion}

In this appendix, we prove that for the XXX model, if $T_{1/2}(u)$ and $T_1(u)$ are polynomials of $u$, then all $T_s(u)$ ($s\geq\frac{3}{2}$,  $2s\in\mathbb{Z}_+$) are automatically polynomials of $u$. Similarly, for the XXZ model, if $T_{1/2}(t)$ and $T_1(t)$ are Laurent polynomials of $t=e^u$, then all $T_s(t)$ are Laurent polynomials.

The fusion relations of the transfer matrices are~\cite{Krichever1997,Miao_2021}
\begin{align}
            T_0^{[(2s-1)+]}\mathbf{T}_{s+1/2}=\mathbf{T}_{1/2}^{[(2s)+]}\mathbf{T}_{s}^{[-]}- T_0^{[(2s+1)+]}\mathbf{T}_{s-1/2}^{[2-]} \; , \label{eq:fusion1}\\
       T_0^{[(2s-1)-]}\mathbf{T}_{s+1/2}=\mathbf{T}_{1/2}^{[(2s)-]}\mathbf{T}_{s}^{[+]}- T_0^{[(2s+1)-]}\mathbf{T}_{s-1/2}^{[2+]} \; ,\label{eq:fusion2}
\end{align}
where $T_0$ is defined as 
\begin{align}
\label{eq:initialT0}
    &T_0 = u^L\,, \qquad \text{for rational case} \;,\\\nonumber
    &T_0 = \sinh^L(u) = \frac{1}{2^L}(t-t^{-1})^L \;, \qquad \text{for trigonometric case}\;.
\end{align}

We prove the result for the XXX case, while the proof for the XXZ case is similar. For $s=1/2$, \eqref{eq:fusion1} and \eqref{eq:fusion2} coincide, yielding
\begin{equation}
    T_0T_1=T_{1/2}^{[+]}T_{1/2}^{[-]}- T_0^{[2+]}T_0^{[2-]} \; .
\end{equation}
For $s=1$, we can multiply $T_0^{[-]}$ on both sides of \eqref{eq:fusion1} and $T_0^{[+]}$ on \eqref{eq:fusion2}, which leads to
\begin{equation}
   \begin{aligned}
        T_0^{[+]}T_0^{[-]}T_{3/2}&=T_0^{[-]}(T_{1/2}^{[2+]}T_{1}^{[-]}- T_0^{[3+]}T_{1/2}^{[2-]})\\&=T_0^{[+]}(T_{1/2}^{[2-]}T_{1}^{[+]}- T_0^{[3-]}T_{1/2}^{[2+]}) \; .
   \end{aligned}
\end{equation}
Therefore,
\begin{equation}
    \frac{T_0^{[+]}}{T_0^{[-]}}(T_{1/2}^{[2-]}T_{1}^{[+]}- T_0^{[3-]}T_{1/2}^{[2+]})=T_{1/2}^{[2+]}T_{1}^{[-]}- T_0^{[3+]}T_{1/2}^{[2-]} \; .
    \label{eq:TT_relation}
\end{equation}
Because the right-hand side of the equation \eqref{eq:TT_relation} is a polynomial, the left-hand side of \eqref{eq:TT_relation} must also be a polynomial. From the definition of $T_0$ in \eqref{eq:initialT0}, it is clear that $T_0^{[-]}$ and $T_0^{[+]}$ have distinct zeros. Therefore, we deduce that the following expression is a polynomial
\begin{equation}
  T_{3/2}=\frac{1}{T_0^{[-]}}(T_{1/2}^{[2-]}T_{1}^{[+]}- T_0^{[3-]}T_{1/2}^{[2+]}) \; .
\end{equation}
Applying this strategy recursively, we have 
\begin{equation}
   \begin{aligned}
        T_0^{[(2s-1)+]}T_0^{[(2s-1)-]}T_{s+1/2}&=T_0^{[(2s-1)-]}\left(T_{1/2}^{[(2s)+]}T_{s}^{[-]}- T_0^{[(2s+1)+]}T_{s-1/2}^{[2-]}\right)\\&=T_0^{[(2s-1)+]}\left(T_{1/2}^{[(2s)-]}T_{s}^{[+]}- T_0^{[(2s+1)-]}T_{s-1/2}^{[2+]}\right) \; .
   \end{aligned}
\end{equation}
This implies that if $T_s$ is a polynomial in $u$, then $T_{s+1/2}$ is a polynomial in $u$ for $s\geq 1$.

\newpage

\section{XXX chain with non-diagonal boundary magnetic field}
\label{app:XXXnondiagonal}

The Hamiltonian of the XXX spin-$\frac{1}{2}$ chain with non-diagonal boundary magnetic field reads~\cite{Wang:2015off}
\begin{equation}
    H=\sum^{L-1}_{k=1}\left(\vec{\sigma}_k\cdot\vec{\sigma}_{k+1}-\mathbb{I}\right)+\frac{1}{\alpha}\left(\sigma_L^z+\xi\sigma_L^x\right)-\frac{1}{\beta}\sigma_1^z \; ,
\end{equation}
where $\mathbb{I}$ is the identity matrix. We define two boundary polynomials as 
\begin{equation}
    f(u)=\left(\sqrt{1+\xi^2}u-\ri\alpha\right)(u+i\beta)=g(-u) \; ,
\end{equation}
and the corresponding $TQ$-relation reads
\begin{equation}
\begin{split}
    -uT_{1/2}Q=&\,g^{[-]}\left(u\Qzz\right)^{[+]}Q^{[2-]}+f^{[+]}\left(u\Qzz\right)^{[-]}Q^{[2+]} \\
    &\,+2\left(1-\sqrt{1+\xi^2}\right)u\left(u\Qzz\right)^{[+]}\left(u\Qzz\right)^{[-]} \; ,
\end{split}
\end{equation}
where $\Qzz(u)=u^{2L}=T_0(u)$ and $Q(u)=\prod^M_{j=1}\left(u-u_j\right)\left(u+u_j\right)$.

When $\xi=0$, $f(u)$ simplifies to $f(u)=(u-\ri\alpha)(u+\ri\beta)$ and the $TQ$-relation becomes homogeneous
\begin{equation}
    -uT_{1/2}Q=g^{[-]}\left(u\Qzz\right)^{[+]}Q^{[2-]}+f^{[+]}\left(u\Qzz\right)^{[-]}Q^{[2+]},
\end{equation}
The $Q$-system has been constructed in~\cite{Nepomechie:2019gqt}.

The fusion relation for the non-diagonal case is~\cite{Miao_2021}
\begin{equation}
    \left(uT_0\right)\left(uT_1\right)=\left(-uT_{1/2}\right)^{[+]}\left(-uT_{1/2}\right)^{[-]}-fg\left(uT_0\right)^{[2+]}\left(uT_0\right)^{[2-]}.
\end{equation}
As described in previous sections, we derive the $Q$-system from the $TQ$ and fusion relation. We define 
\begin{align}
    &uQ_{1,1} = f Q^{[+]}_{1,0} - g Q^{[-]}_{1,0} \; , \\
  &uQ_{1,2} = f^{[-]} Q^{[+]}_{1,1} - g^{[+]} Q^{[-]}_{1,1} \; .
\end{align}
The corresponding $QQ$-relations are as follows,
\begin{align}
    &u\Qzo\Qoz=\Qzz^{[+]} \Qoo^{[-]} - \Qzz^{[-]} \Qoo^{[+]}-2\left(1-\sqrt{1+\xi^2}\right)u\Qzz^{[+]}\Qzz^{[-]}\;,\\
    &u\Qzt\Qoo=\Qzo^{[+]} \Qot^{[-]} - \Qzo^{[-]} \Qot^{[+]} -2\left(1-\sqrt{1+\xi^2}\right)\left(fQ_{0,0}^{[2+]}Q_{0,1}^{[-]}-gQ_{0,0}^{[2-]}Q_{0,1}^{[+]}\right)\;.
\end{align}
The $Q$-system can be solved in the same way as in the main text, which yields all physical solutions.

\section{Details for derivations}
\label{app:detailsforderivations}

In this appendix, we provide more details for the derivations in the main text.

\subsection{Details for the anti-diagonal twisted case}
\label{app:QQ_relation_2nd_box}

In this subsection, we give details for the $QQ$-relation \eqref{eq:Qsystem2ndAD}. We highlight the general twist parameter $\alpha$ and $\beta$ in this section for better clarity.

We first need to derive a $TQ$-like relation for the higher spin transfer matrix $T_1$. Multiplying both sides of \eqref{eq:fusionAD1} with $\Qoo$
\begin{align}
    \label{eq:product_T0T1Q11}
        T_0 T_1 \Qoo =&\, T^{[+]}_{1/2} T^{[-]}_{1/2} \Qoo + \textcolor{blue}{\alpha \beta} T^{[2+]}_0 T^{[2-]}_0 \Qoo \\\nonumber
        =&\, T^{[-]}_{1/2} t^{[-]} T^{[+]}_{1/2} Q^{[+]}_{1,0} - T^{[+]}_{1/2} (t^{-1})^{[+]} T^{[-]}_{1/2} Q^{[-]}_{1,0}\\\nonumber
        &\, + \textcolor{blue}{\alpha \beta} T^{[2+]}_0 T^{[2-]}_0 t^{[-]} Q^{[+]}_{1,0} - \textcolor{blue}{\alpha \beta} T^{[2+]}_0 T^{[2-]}_0 (t^{-1})^{[+]} Q^{[-]}_{1,0} \; ,
\end{align}
where we have used the definition of $\Qoo$ \eqref{eq:Q11AD}. Using the $TQ$-relation in \eqref{eq:TQAD}, we obtain
\begin{equation}
    \begin{aligned}
          T_0 T_1 \Qoo =& T^{[-]}_{1/2} t^{[-]} q^{\frac{1}{2}}\left(\textcolor{blue}{\alpha} t^{[+]} Q^{[3+]}_{1,0} T_0 - \textcolor{blue}{\beta} (t^{-1})^{[+]} Q^{[-]}_{1,0} T_0^{[2+]} - \textcolor{blue}{\alpha \beta} (c(t))^{[+]} T_0^{[2+]} T_0\right) \\
  &- T^{[+]}_{1/2} (t^{-1})^{[+]} q^{\frac{1}{2}}\left(\textcolor{blue}{\alpha} t^{[-]} Q^{[+]}_{1,0} T_0^{[2-]} - \textcolor{blue}{\beta} (t^{-1})^{[-]} Q^{[3-]}_{1,0} T_0 - \textcolor{blue}{\alpha \beta} (c(t))^{[-]} T_0 T_0^{[2-]}\right) \\
  &+ \textcolor{blue}{\alpha \beta} T^{[2+]}_0 T^{[2-]}_0 t^{[-]} Q^{[+]}_{1,0} - \textcolor{blue}{\alpha \beta} T^{[2+]}_0 T^{[2-]}_0 (t^{-1})^{[+]} Q^{[-]}_{1,0} \; .
    \end{aligned}
\end{equation}
Use $TQ$-relation \eqref{eq:TQAD} once more for $T_{1/2}$ and divide both sides by $T_0$, we obtain
 \begin{align}
  T_1 \Qoo =&\, \textcolor{blue}{\alpha} t t^{[+]} T^{[-]}_{1/2} Q^{[3+]}_{1,0} + \textcolor{blue}{\beta^2} (t^{-1})^{[-]} Q^{[3-]}_{1,0} T_0^{[2+]}\\\nonumber
   &\,- \textcolor{blue}{\alpha^2} t^{[+]} Q^{[3+]}_{1,0} T_0^{[2-]} + \textcolor{blue}{\beta} (t^{-1}) (t^{-1})^{[-]} T^{[+]}_{1/2} Q^{[3-]}_{1,0} \\\nonumber
  &\,+ \textcolor{blue}{\alpha \beta^2} (c(t))^{[-]} T_0^{[2+]} T_0^{[2-]} + \textcolor{blue}{\alpha^2 \beta} (c(t))^{[+]} T_0^{[2+]} T_0^{[2-]}  - \textcolor{blue}{\alpha \beta} t (c(t))^{[+]} T^{[-]}_{1/2} T_0^{[2+]}\\\nonumber
  &\, + \textcolor{blue}{\alpha \beta} t^{-1} (c(t))^{[-]}  T^{[+]}_{1/2} T_0^{[2-]} \; . \label{eq:TQ for T1}
\end{align}
We expect a similar $QQ$-relation involving $Q_{0,2}$, which takes the form 
\begin{equation}
  \textcolor{blue}{\beta} \Qzo^{[+]} \Qot^{[-]} + \textcolor{blue}{\alpha} \Qzo^{[-]} \Qot^{[+]} {=} \Qzt \Qoo+\texttt{inhomogeneous terms} \; . \label{eq:conj_Q-system_2nd_box}
\end{equation}
To calculate the left hand side, we need the definition \eqref{eq:Q12AD}, which gives 
\begin{align}
  \textcolor{blue}{\beta} \Qzo^{[+]} \Qot^{[-]} + \textcolor{blue}{\alpha} \Qzo^{[-]} \Qot^{[+]} = &\,\textcolor{blue}{\beta} t^{[-]} \Qzo^{[+]} Q_{1,1} - \textcolor{blue}{\beta} (t^{-1})^{[-]} \Qzo^{[+]} Q^{[2-]}_{1,1}\\\nonumber
  &\, + \textcolor{blue}{\alpha} t^{[+]} \Qzo^{[-]} Q^{[2+]}_{1,1} - \textcolor{blue}{\alpha} (t^{-1})^{[+]} \Qzo^{[-]} Q_{1,1} \; .
\end{align}

Combining the definitions \eqref{eq:Q11AD} and \eqref{eq:Q01AD}, we have
\begin{equation}
    \begin{aligned}
  &\textcolor{blue}{\alpha} t^{[+]} \Qzo^{[-]} Q^{[2+]}_{1,1} - \textcolor{blue}{\beta} (t^{-1})^{[-]} \Qzo^{[+]} Q^{[2-]}_{1,1} \\
  =& \textcolor{blue}{\alpha} t^{[+]} \left(q^{-\frac{1}{2}} T^{[-]}_{1/2} + \textcolor{blue}{\beta} t^{[3-]} T_0 - \textcolor{blue}{\alpha} (t^{-1})^{[+]} T_0^{[2-]}\right) \left(t^{[+]} Q^{[3+]}_{1,0} - (t^{-1})^{[3+]} Q^{[+]}_{1,0}\right) \\
  &+ \textcolor{blue}{\beta} (t^{-1})^{[-]} \left(q^{-\frac{1}{2}} T^{[+]}_{1/2} + \textcolor{blue}{\beta} t^{-} T_0^{[2+]} - \textcolor{blue}{\alpha} (t^{-1})^{[3+]} T_0\right) \left((t^{-1})^{[-]} Q^{[3-]}_{1,0} - t^{[3-]} Q^{[-]}_{1,0}\right) \\
  =& \textcolor{blue}{\alpha} t t^{[+]} T^{[-]}_{1/2} Q^{[3+]}_{1,0} + \textcolor{blue}{\beta^2} (t^{-1})^{[-]} T_0^{[2+]} Q^{[3-]}_{1,0} - \textcolor{blue}{\alpha^2} t^{[+]} T_0^{[2-]} Q^{[3+]}_{1,0} + \textcolor{blue}{\beta} (t^{-1}) (t^{-1})^{[-]} T^{[+]}_{1/2} Q^{[3-]}_{1,0} \\
  &+ \textcolor{blue}{\alpha \beta} t^2 t^{[-]} T_0 Q^{[3+]}_{1,0} - \textcolor{blue}{\beta^2} t^{[3-]} T_0^{[2+]} Q^{[-]}_{1,0} + \textcolor{blue}{\alpha^2} (t^{-1})^{[3+]} T_0^{[2-]} Q^{[+]}_{1,0} - \textcolor{blue}{\alpha \beta} (t^{-2}) (t^{-1})^{[+]} T_0 Q^{[3-]}_{1,0} \\
  &- \textcolor{blue}{\alpha} t (t^{-1})^{[3+]} T^{[-]}_{1/2} Q^{[+]}_{1,0} - \textcolor{blue}{\alpha \beta} t^{[5-]} T_0 Q^{[+]}_{1,0} - \textcolor{blue}{\beta} (t^{-1}) t^{[3-]} T^{[+]}_{1/2} Q^{[-]}_{1,0} + \textcolor{blue}{\alpha \beta} (t^{-1})^{[5+]} T_0 Q^{[-]}_{1,0} \; . \label{eq:2_terms_Q02}
\end{aligned}
\end{equation}
Note that in the last step, the first line can be replaced by $T_1\Qoo$ subtracting an inhomogeneous term that includes all $c(t)$,
\begin{equation}
\begin{aligned}
    &\textcolor{blue}{\alpha} t t^{[+]} T^{[-]}_{1/2} Q^{[3+]}_{1,0} + \textcolor{blue}{\beta^2} (t^{-1})^{[-]} T_0^{[2+]} Q^{[3-]}_{1,0} - \textcolor{blue}{\alpha^2} t^{[+]} T_0^{[2-]} Q^{[3+]}_{1,0} + \textcolor{blue}{\beta} (t^{-1}) (t^{-1})^{[-]} T^{[+]}_{1/2} Q^{[3-]}_{1,0} \\
    =& T_1\Qoo - f(c(t)) \;,
    \label{eq:result_of_1st_line}
\end{aligned}
\end{equation}
where $f(c(t))$ is defined as 
\begin{align}
 f(c(t)) =&\,\textcolor{blue}{\alpha \beta^2} (c(t))^{[-]} T_0^{[2+]} T_0^{[2-]} + \textcolor{blue}{\alpha^2 \beta} (c(t))^{[+]} T_0^{[2+]} T_0^{[2-]}\\\nonumber
 &\, - \textcolor{blue}{\alpha \beta} t (c(t))^{[+]} T^{[-]}_{1/2} T_0^{[2+]} + \textcolor{blue}{\alpha \beta} t^{-1} (c(t))^{[-]}  T^{[+]}_{1/2} T_0^{[2-]} \; .
\end{align}
Using
\begin{equation}
  T_{1/2} = q^{\frac{1}{2}} \left(\Qzo - \textcolor{blue}{\beta} t^{[2-]} T_0^{[+]} + \textcolor{blue}{\alpha} (t^{-1})^{[2+]} T_0^{[-]}\right)\;,
\end{equation}
we express $f(c(t))$ in terms of $Q$-functions~\footnote{Note that $T_0 = \Qzz$.}
 \begin{align}
  f(c(t)) =&\, \textcolor{blue}{\alpha \beta^2} (c(t))^{[-]} T_0^{[2+]} T_0^{[2-]} + \textcolor{blue}{\alpha^2 \beta} (c(t))^{[+]} T_0^{[2+]} T_0^{[2-]}\\\nonumber
    &\,- \textcolor{blue}{\alpha \beta} t (c(t))^{[+]} T^{[-]}_{1/2} T_0^{[2+]} + \textcolor{blue}{\alpha \beta} t^{-1} (c(t))^{[-]}  T^{[+]}_{1/2} T_0^{[2-]} \\\nonumber
  =&\, \textcolor{blue}{\alpha \beta} (t^{-1})^{[-]} (c(t))^{[-]} \Qzo^{[+]} T_0^{[2-]} - \textcolor{blue}{\alpha \beta} t^{[+]} (c(t))^{[+]} \Qzo^{[-]} T_0^{[2+]} \\\nonumber
  \label{eq:def_f(c(t))}
  &+ \textcolor{blue}{\alpha \beta^2} t t^{[2-]} (c(t))^{[+]} T_0 T_0^{[2+]} + \textcolor{blue}{\alpha^2 \beta} (t^{-1}) (t^{-1})^{[2+]} (c(t))^{[-]} T_0 T_0^{[2-]} \; ,
\end{align}
After some manipulation to $\Qoz^{[\pm]}$ in \eqref{eq:2_terms_Q02}, we have 
\begin{equation}
\begin{aligned}
    \textcolor{blue}{\alpha \beta} t^2 t^{[-]} T_0 Q^{[3+]}_{1,0} - \textcolor{blue}{\beta^2} t^{[3-]} T_0^{[2+]} Q^{[-]}_{1,0} =& \textcolor{blue}{\beta} t t^{[2-]} \left(\textcolor{blue}{\alpha} t T_0^- \Qoz^{[2+]} - \textcolor{blue}{\beta} t^{-1} T_0^+ \Qoz^{[2-]} \right)^{[+]} \\
    =& \textcolor{blue}{\beta} t t^{[2-]} \left(T_{1/2} Q_{1,0} q^{-\frac{1}{2}} + \textcolor{blue}{\alpha \beta} c(t) T_0^{[+]} T_0^{[-]}\right)^{[+]} \; .
\end{aligned}
\end{equation}
Similarly, 
\begin{align}
    &\textcolor{blue}{\alpha^2} (t^{-1})^{[3+]} T_0^{[2-]} Q^{[+]}_{1,0} - \textcolor{blue}{\alpha \beta} (t^{-2}) (t^{-1})^{[+]} T_0 Q^{[3-]}_{1,0}\\\nonumber
     =&\, \textcolor{blue}{\alpha} t^{-1} (t^{-1})^{[2+]} \left(\textcolor{blue}{\alpha} t T_0^- \Qoz^{[2+]} - \textcolor{blue}{\beta} t^{-1} T_0^+ \Qoz^{[2-]} \right)^{[-]} \\\nonumber
    =&\, \textcolor{blue}{\alpha} t^{-1} (t^{-1})^{[2+]} \left(T_{1/2} Q_{1,0} q^{-\frac{1}{2}} + \textcolor{blue}{\alpha \beta} c(t) T_0^{[+]} T_0^{[-]}\right)^{[-]} \; .
\end{align}
Recall the definition of $\Qoo$ \eqref{eq:Q11AD} and combine with \eqref{eq:result_of_1st_line}, there is
\begin{equation}
    \begin{aligned}
  &\textcolor{blue}{\alpha} t^{[+]} \Qzo^{[-]} Q^{[2+]}_{1,1} - \textcolor{blue}{\beta} (t^{-1})^{-} \Qzo^{[+]} Q^{[2-]}_{1,1} \\
  =& T_1 \Qoo + \textcolor{blue}{\beta}t^{[2-]} T_{1/2}^{[+]} \Qoo - \textcolor{blue}{\alpha} (t^{-1})^{[2+]} T_{1/2}^{[-]} \Qoo - \textcolor{blue}{\alpha \beta} q^{-2} T_0 \Qoo \\
  &+ \textcolor{blue}{\alpha^2 \beta} (t^{-1}) (t^{-1})^{[2+]} (c(t))^{[-]} T_0 T_0^{[2-]} + \textcolor{blue}{\alpha \beta^2} t t^{[2-]} (c(t))^{[+]} T_0^{[2+]} T_0 - f(c(t)) \; .
\end{aligned}
\end{equation}
As conjectured in \eqref{eq:conj_Q-system_2nd_box}, we define $\Qzt$ as 
\begin{equation}
  \Qzt = T_1 + \textcolor{blue}{\beta} t^{[2-]} T_{1/2}^{[+]} - \textcolor{blue}{\alpha} (t^{-1})^{[2+]} T_{1/2}^{[-]} - \textcolor{blue}{\alpha \beta} q^{-2} T_0 + \textcolor{blue}{\beta} t^{[-]} \Qzo^{[+]} - \textcolor{blue}{\alpha} (t^{-1})^{[+]} \Qzo^{[-]}, \label{eq:def_Q02_containing_Q01}
\end{equation}
and define a new inhomogeneous term as 
\begin{equation}
    \begin{aligned}
          F(t) =&\, \textcolor{blue}{\alpha \beta^2} t t^{[2-]} (c(t))^{[+]} T_0 T_0^{[2+]} + \textcolor{blue}{\alpha^2 \beta} (t^{-1}) (t^{-1})^{[2+]} (c(t))^{[-]} T_0^{[2-]} T_0 - f(c(t)) \\
  =&\, \textcolor{blue}{\alpha \beta} t^{[+]} (c(t))^{[+]} \Qzo^{[-]} T_0^{[2+]} - \textcolor{blue}{\alpha \beta} (t^{-1})^{[-]} (c(t))^{[-]} \Qzo^{[+]} T_0^{[2-]} \; .
    \end{aligned}
\end{equation}
Hence, we obtain a compact $QQ$-relation
\begin{equation}
  \Qzt \Qoo = \textcolor{blue}{\beta} \Qzo^{[+]} \Qot^{[-]} + \textcolor{blue}{\alpha} \Qzo^{[-]} \Qot^{[+]} - F(t).
\end{equation}

\subsection{Details for the open case}
\label{app:derivation_open_case}

In this subsection, we give details for $T_1Q_{1,1}$-relation~\eqref{eq:open_T1Q11} and $QQ$-relation~\eqref{eq:Q01Q12_final}.

\paragraph{$T_1Q_{1,1}$ relation.} 
We start by considering the shifted $TQ$-relation of \eqref{eq:open_XXZ_TQ}, 
\begin{equation}
    \begin{split}
 &U^{[+]} T^{[+]}_\frac{1}{2} \Qoz^{[+]} = - 4 \left( U^{[2+]} f T_0^{[2+]} \Qoz^{-} + U g^{[2+]} T_0 Q^{[3+]}  \right)+ 2 x T_0^{[2+]} T_0 U^{[2+]} U^{[+]} U \; , \\
   &U^{[-]} T_\frac{1}{2}^{[-]} \Qoz^{[-]} = - 4 \left( U f^{[2-]} T_0 Q^{[3-]} + U^{[2-]} g T_0^{[2-]} Q^{[+]}  \right)  + 2 x T_0 T_0^{[2-]} U U^{[-]} U^{[2-]}\; .
   \end{split}
\label{eq:shift_open_XXZ_TQ}
\end{equation}
Applying \eqref{eq:shift_open_XXZ_TQ} to \eqref{eq:UT0UT1Q11}, we have
\begin{align}
    &(U T_0) (U T_1)\Qoo\label{eq:open_T0T1Q11}=\\&\frac{U^{[-]}T_\frac{1}{2}^{[-]}}{U}g\left(- 4 \left[ U^{[2+]} f T_0^{[2+]} \Qoz^{-} + U g^{[2+]} T_0 Q^{[3+]}  \right]
     + 2 x T_0^{[2+]} T_0 U^{[2+]} U^{[+]} U\right)\label{eq:first_term_open_T0T1Q11}\\&
     -\frac{U^{[+]}T_\frac{1}{2}^{[+]}}{U}f\left(- 4 \left[ U f^{[2-]} T_0 Q^{[3-]} + U^{[2-]} g T_0^{[2-]} Q^{[+]}  \right]
     + 2 x T_0 T_0^{[2-]} U U^{[-]} U^{[2-]}\right)\label{eq:second_term_open_T0T1Q11}\\&
     -16 f g (U T_0)^{[2+]} (U T_0)^{[2-]}\frac{1}{U}\left(g Q^{[+]}_{1,0} - f Q^{[-]}_{1,0}\right)\label{eq:third term of open T0T1Q11} .
\end{align}

For the first part \eqref{eq:first_term_open_T0T1Q11}, applying \eqref{eq:shift_open_XXZ_TQ} again, we obtain
\begin{equation}
    \begin{aligned}
    &\frac{U^{[-]}T_\frac{1}{2}^{[-]}}{U}g\left(- 4 \left[ U^{[2+]} f T_0^{[2+]} \Qoz^{-} + U g^{[2+]} T_0 Q^{[3+]}  \right]
     + 2 x T_0^{[2+]} T_0 U^{[2+]} U^{[+]} U\right)\\
     &=16ff^{[2-]}gU^{[2+]}T_0^{[2+]}\Qoz^{[3-]}T_0+\frac{16fg^2U^{[2+]}U^{[2-]}}{U}T_0^{[2-]}T_0^{[2+]}\Qoz^{[+]}\\&-8xfgU^{[-]}U^{[2-]}U^{[2+]}T_0^{[2-]}T_0^{[2+]}T_0\\&-4gg^{[2+]}U^{[-]}T_{1/2}^{[-]}\Qoz^{[3+]}T_0+2xgU^{[-]}U^{[2+]}U^{[+]}T_0T_0^{[2+]}T_\frac{1}{2}^{[-]} \; .
    \end{aligned}
\end{equation}
Similarly, the second part \eqref{eq:second_term_open_T0T1Q11} becomes
\begin{equation}
    \begin{aligned}
    &-\frac{U^{[+]}T_\frac{1}{2}^{[+]}}{U}f\left[- 4 \left( U f^{[2-]} T_0 Q^{[3-]} + U^{[2-]} g T_0^{[2-]} Q^{[+]}  \right)
     + 2 x T_0 T_0^{[2-]} U U^{[-]} U^{[2-]}\right]\\
     &=-16fgg^{[2+]}U^{[2-]}T_0^{[2-]}\Qoz^{[3+]}T_0-16f^2g\frac{U^{[2-]}U^{[2+]}}{U}T_0^{[2-]}T_0^{[2+]}\Qoz^{[-]}\\&+8xfgU^{[2-]}U^{[2+]}U^{[+]}T_0^{[2-]}T_0^{[2+]}T_0\\&+4ff^{[2-]}U^{[+]}T_\frac{1}{2}^{[+]}\Qoz^{[3-]}T_0-2xfU^{[+]}U^{[-]}U^{[2-]}T_\frac{1}{2}^{[+]}T_0^{[2-]}T_0 \; .
     \end{aligned}
\end{equation}
Combining both parts above, we find that terms without $T_0$ are canceled, proving \eqref{eq:open_T1Q11}.

\paragraph{$QQ$-relation} 
We need to perform the calculation of the expression enclosed in the first pair of brackets in \eqref{eq:Q01Q12}. From \eqref{eq:def_open_Qzo}, we have
\begin{align}
     &\frac{g\Qzo^{[-]}\Qoo^{[2+]}}{U^{[+]}}+\frac{f\Qzo^{[+]}\Qoo^{[2-]}}{U^{[-]}}\\&=\frac{gg^{[2+]}T_\frac{1}{2}^{[-]}\Qoz^{[3+]}}{4U^{[2-]}UU^{[+]}U^{[2+]}}+\frac{fgg^{[2+]}T_0^{[2-]}\Qoz^{[3+]}}{U^{[-]}UU^{[+]}U^{[2+]}}-\frac{ff^{[2-]}T_\frac{1}{2}^{[+]}\Qoz^{[3-]}}{4U^{[2-]}U^{[-]}UU^{[2+]}}-\frac{fgf^{[2-]}T_0^{[2+]}\Qoz^{[3-]}}{U^{[2-]}U^{[-]}UU^{[+]}}\label{eq:3.34}\\&+\frac{gg^{[2+]}g^{[2-]}T_0\Qoz^{[3+]}}{U^{[2-]}U^{[-]}U^{[+]}U^{[2+]}}+\frac{fgg^{[2-]}T_0^{[2+]}\Qoz^{-}}{U^{[2-]}U^{[-]}UU^{[+]}}-\frac{fgf^{[2+]}T_0^{[2-]}\Qoz^{[+]}}{U^{-}UU^{[+]}U^{[2+]}}-\frac{ff^{[2+]}f^{[2-]}T_0\Qoz^{[3-]}}{U^{[2-]}U^{[-]}U^{[+]}U^{[2+]}}\label{eq:3.35}\\&-\frac{gf^{[2+]}T_{1/2}^{[-]}\Qoz^{[+]}}{4U^{[2-]}UU^{[+]}U^{[2+]}}-\frac{gg^{[2-]}f^{[2+]}T_0\Qoz^{[+]}}{U^{[2-]}U^{[-]}U^{[+]}U^{[2+]}}+\frac{fg^{[2-]}T_{1/2}^{[+]}\Qoz^{[-]}}{4U^{[2-]}UUU^{[2+]}}+\frac{ff^{[2+]}g^{[2-]}T_0\Qoz^{[-]}}{U^{[2-]}U^{[-]}U^{[+]}U^{[2+]}} \; . 
     \label{eq:3.36}
\end{align}

For the first four terms of \eqref{eq:3.34}, we use \eqref{eq:open_T1Q11} to simplify
\begin{equation}
    \begin{aligned}
    &\frac{gg^{[2+]}T_\frac{1}{2}^{[-]}\Qoz^{[3+]}}{4U^{[2-]}UU^{[+]}U^{[2+]}}+\frac{fgg^{[2+]}T_0^{[2-]}\Qoz^{[3+]}}{U^{[-]}UU^{[+]}U^{[2+]}}-\frac{ff^{[2-]}T_\frac{1}{2}^{[+]}\Qoz^{[3-]}}{4U^{[2-]}U^{[-]}UU^{[2+]}}-\frac{fgf^{[2-]}T_0^{[2+]}\Qoz^{[3-]}}{U^{[2-]}U^{[-]}UU^{[+]}}\\
    &=-\frac{U^2T_1Q_1-S(x)}{16U^{[2-]}U^{[-]}UU^{[+]}U^{[2+]}} \; .
    \end{aligned}
\end{equation}

To simplify the first two terms in \eqref{eq:3.35}, we use the shifted $TQ$ relation \eqref{eq:shift_open_XXZ_TQ},
\begin{equation}
    \begin{aligned}
        \frac{gg^{[2+]}g^{[2-]}T_0\Qoz^{[3+]}}{U^{[2-]}U^{[-]}U^{[+]}U^{[2+]}}+\frac{fgg^{[2-]}T_0^{[2+]}\Qoz^{-}}{U^{[2-]}U^{[-]}UU^{[+]}}
        &=-\frac{gg^{[2-]}\left(U^{[+]}T_\frac{1}{2}^{[+]}\Qoz^{[+]}-2xT_0^{[2+]}T_0UU^{[+]}U^{[2+]}\right)}{4U^{[2-]}U^{[-]}UU^{[+]}U^{[2+]}}\; .
    \end{aligned}
\end{equation}
Similarly, the last two terms in \eqref{eq:3.35} become
\begin{equation}
    \begin{aligned}
        -\frac{fgf^{[2+]}T_0^{[2-]}\Qoz^{[+]}}{U^{-}UU^{[+]}U^{[2+]}}-\frac{ff^{[2+]}f^{[2-]}T_0\Qoz^{[3-]}}{U^{[2-]}U^{[-]}U^{[+]}U^{[2+]}}=\frac{ff^{[2+]}\left(U^{[-]}T_\frac{1}{2}^{[-]}\Qoz^{[-]}-2xT_0^{[2-]}T_0U^{[2-]}U^{[-]}U\right)}{4U^{[2-]}U^{[-]}UU^{[+]}U^{[2+]}} \; .
    \end{aligned}
\end{equation}

Combining all the results above and using \eqref{eq:def_open_Q11}, we finally have
\begin{align}
    \label{eq:left_bracket}
       & -\left[\frac{g\Qzo^{[-]}\Qoo^{[2+]}}{U^{[+]}}+\frac{f\Qzo^{[+]}\Qoo^{[2-]}}{U^{[-]}}\right]\\\nonumber
       =&\,\frac{U^2T_1Q_1-S(x)}{16U^{[2-]}U^{[-]}UU^{[+]}U^{[2+]}}+\frac{g^{[2-]}T_\frac{1}{2}^{[+]}\Qoo}{4U^{[2-]}U^{[-]}U^{[2+]}}+\frac{f^{[2+]}T_\frac{1}{2}^{[-]}\Qoo}{4U^{[2-]}U^{[+]}U^{[2+]}}\\\nonumber
       &\,-\frac{xgg^{[2-]}T_0T_0^{[2+]}}{2U^{[2-]}U^{[-]}}+\frac{xff^{[2+]}T_0T_0^{[2-]}}{2U^{[+]}U^{[2+]}}+\frac{g^{[2-]}f^{[2+]}T_0U\Qoo}{U^{[2-]}U^{[-]}U^{[+]}U^{[2+]}}\; ,
\end{align}
which leads to \eqref{eq:Q01Q12_final}.

\bibliographystyle{JHEP}
\bibliography{biblio.bib}

\end{document}